\documentclass[12pt]{article}
\usepackage[reqno]{amsmath}
\usepackage{epsfig}
\usepackage{array}
\usepackage{float}

\usepackage{a4}
\usepackage{epsfig}
\usepackage{a4wide}
\usepackage{wasysym}

\def\Jo#1#2#3#4{{\it #1} {\bf #2}, #3 (#4)}

\def\PRL{Phys. Rev. Lett.}

\def\ra{\rightarrow}
\def\be{\begin{equation}}
\def\ee{\end{equation}}
\def\gs{\mathrel{
   \rlap{\raise 0.511ex \hbox{$>$}}{\lower 0.511ex \hbox{$\sim$}}}}
\def\ls{\mathrel{
   \rlap{\raise 0.511ex \hbox{$<$}}{\lower 0.511ex \hbox{$\sim$}}}}
\newcommand{\obb}{0\mbox{$\nu\beta\beta$}}
\newcommand{\onbb}{neutrinoless double beta decay}

\newcommand{\ba}{\begin{array}{c}}
\newcommand{\baz}{\begin{array}{cc}}
\newcommand{\bad}{\begin{array}{ccc}}
\newcommand{\bav}{\begin{array}{cccc}}
\newcommand{\bea}{\begin{equation} \begin{array}{c}}
\newcommand{\eea}{ \end{array} \end{equation}}
\newcommand{\ea}{\end{array}}
\newcommand{\D}{\displaystyle}
\newcommand{\dms}{\mbox{$\Delta m^2_{\odot}$}}
\newcommand{\dma}{\mbox{$\Delta m^2_{\rm A}$}}
\newcommand{\meff}{\mbox{$\langle m \rangle$}}

\def\ra{\rightarrow}

\def\gtap{\mathrel{
   \rlap{\raise 0.511ex \hbox{$>$}}{\lower 0.511ex \hbox{$\sim$}}}}
\def\ltap{\mathrel{
   \rlap{\raise 0.511ex \hbox{$<$}}{\lower 0.511ex \hbox{$\sim$}}}}

\newcommand{\PMNS}{\mbox{$ U_{\rm PMNS}$}}

\begin{document}

\title{
\hfill {\small SISSA 83/2003/EP}\\
\vspace{-0.3cm}
\hfill {\small hep-ph/0309249} \\ 
\vskip 0.5cm
\bf A Parametrization for the Neutrino Mixing Matrix
}
\author{
Werner Rodejohann\\ \\
{\normalsize \it Scuola Internazionale Superiore di Studi Avanzati,
I-34014 Trieste, Italy}\\
{ \normalsize and} \\
{\normalsize \it Istituto Nazionale di Fisica Nucleare,
Sezione di Trieste, I-34014 Trieste, Italy}\\
}
\date{}
\maketitle
\thispagestyle{empty}
\vspace{-0.8cm}
\begin{abstract}
\noindent We propose a flexible and model independent 
parametrization of the neutrino mixing matrix, which takes 
advantage of the fact that there are up to three small quantities 
in neutrino mixing phenomenology: 
(i) the deviation from maximal mixing of solar neutrinos, 
(ii) the mixing matrix element $U_{e3}$ and 
(iii) the deviation from maximal mixing of atmospheric neutrinos. 
It is possible to quantify those three observations 
with a parameter $\lambda \sim 0.2$, which appears at least linearly in 
all elements of the mixing matrix. 
The limit $\lambda \rightarrow 0$ corresponds to exact bimaximal 
mixing. 
Present and future experiments can be used to pin down the power of 
$\lambda$ required to usefully describe the observed phenomenology. 
Observing that the ratio of the two measured 
mass squared differences is 
roughly $\lambda^2$ allows to further 
study the structure of the Majorana mass matrix. 
We comment on the implications of this 
parametrization for neutrinoless double beta decay and on the 
oscillation probabilities in long--baseline experiments.

\end{abstract}

\newpage

\section{\label{sec:intro}Introduction}
Neutrino physics has made impressive progress in recent 
years \cite{reviews}. In particular, the structure of the 
neutrino mixing matrix has been identified to a reasonable precision. 
The final step for the determination of its structure has come from the 
KamLAND experiment \cite{KamLAND}, which confirmed the large mixing angle 
solution for the solar neutrino problem, after a preference for this 
parameter space was already implied by the data of the various 
solar neutrino experiments \cite{sol}. 
Very recently, the SNO salt phase data \cite{SNOII} rejected maximal solar 
mixing by more than $5 \sigma$ \cite{SNOII,SNOana1,SNOana2}.  
The (almost) maximal mixing of atmospheric neutrinos has been found by the 
SuperKamiokande experiment \cite{SKatm} and confirmed by the 
K2K collaboration \cite{K2K}. Finally, the 
presence of a small if not zero angle was 
implied by reactor experiments \cite{reactor}. 
In the present paper we wish to propose a parametrization of the neutrino 
mixing matrix in terms of a small parameter $\lambda$, whose magnitude is 
interestingly around 0.2, i.e., close to the Wolfenstein parameter used to 
parametrize the CKM matrix \cite{wolf}. 
In any parametrization of the neutrino mixing 
matrix (for earlier attempts, see \cite{kaus, zing, zee}), 
it is convenient to start from a reference matrix and describe 
deviations from it.  
Our reference matrix is the one corresponding to exact bimaximal 
neutrino mixing. The parameter $\lambda$ describes the up to three 
small deviations from this mixing scheme, namely the 
deviation from maximal mixing of solar neutrinos, the 
deviation from zero $U_{e3}$ and the deviation from maximal mixing of 
atmospheric neutrinos. The magnitude of $\lambda$ is defined by the 
observed non--maximality of solar neutrino mixing \cite{SNOII,SNOana1,SNOana2} 
and future precision experiments can be used to 
pin down the power of $\lambda$ to usefully describe the other two 
deviations. 
In addition, the ratio of the mass squared 
differences governing solar and atmospheric neutrino oscillations is 
given by $\lambda^2$, so that it is possible to 
analyze also the structure of the neutrino 
mass matrix (provided neutrinos are Majorana particles). 
We also analyze the oscillation probabilities for long 
baseline experiments and the effective mass as measured in 
\onbb{} within our parametrization.\\ 

\noindent The paper is organized as follows: 
In Section \ref{sec:vs} we describe the neutrino mixing parameters 
as implied 
by current data and outline the idea of our parametrization. 
Then, in Section \ref{sec:U} we give the form of the mixing matrix 
for various special cases of the parametrization and analyze in Section 
\ref{sec:mnu} the form of the neutrino mass matrix. 
In Section \ref{sec:app} we apply our parametrization to the 
effective mass as measured in \onbb{} 
and to long--baseline oscillation experiments. 
We conclude in Section \ref{sec:concl}.

\section{\label{sec:vs}Quark versus Lepton Mixing}
The Wolfenstein parametrization \cite{wolf} 
of the CKM matrix uses the fact that 
the quark mixing is very small, i.e., the mixing matrix is 
quasi the unit matrix with only 
small corrections on the off--diagonal entries. 
In terms of mixing angles, a hierarchy of the form 
$\theta_{12} \sim 0.1 > \theta_{23} \sim 0.01 > \theta_{13} \sim 0.001$ 
is observed. This has been used by Wolfenstein to 
introduce an expansion parameter $\lambda$ 
describing the mixing between $u$ and $s$ quarks. The observation that 
$c-b$ ($u-b$) mixing is roughly one (two) 
orders of magnitude suppressed then leads to   
\be \label{eq:wolf} 
V_{\rm CKM} \simeq 
\left( 
\bad 
1 - \frac{1}{2} \, \lambda^2 & \lambda & A \, \lambda^3 \, 
(\rho - i \eta) \\[0.3cm]
-\lambda & 1 - \frac{1}{2} \, \lambda^2 & A \, \lambda^2 \\[0.3cm]
A \, \lambda^3 \, (1 - \rho + i \eta) & -A \, \lambda^2 & 1 
\ea
\right) + {\cal{O}}(\lambda^4)~.
\ee
Of course, $\lambda$ corresponds to the Cabibbo angle 
$\sin \theta_C \simeq 0.22$, whereas the other parameters are about \cite{PDG} 
$A \simeq 0.83$, $\rho \simeq 0.23$ and $\eta \simeq 0.36$. 
The latter parameter describes the $CP$ violation in the quark sector; 
all such effects have to be proportional to \cite{JCP}
\be
J_{CP} = {\rm Im} \{ V_{ud} \, V_{cb} \, V_{ub}^\ast \, V_{cd}^\ast \} 
 \simeq -A^2 \, \lambda^6 \, \eta \sim -3 \cdot 10^{-5}~.
\ee
Therefore, $CP$ violation in the quark sector is a small effect.

\subsection{Neutrino Mixing}
The neutrino oscillation data 
can consistently be described within a
3--neutrino mixing scheme with massive neutrinos, in which the 
flavor states $\nu_\alpha$ ($\alpha = e, \mu, \tau$) are mixed 
with the mass states $\nu_i$ ($i = 1,2,3$) via 
$U_{\rm PMNS}$, the unitary 
Pontecorvo--Maki--Nagakawa--Sakata 
\cite{PMNS} lepton 
mixing matrix. It can be parametrized as 
\bea \label{eq:Upara}
\PMNS = \left( \bad 
c_{12} c_{13} & s_{12} c_{13} & s_{13}  \\[0.2cm] 
-s_{12} c_{23} - c_{12} s_{23} s_{13} e^{i \delta} 
& c_{12} c_{23} - s_{12} s_{23} s_{13} e^{i \delta} 
& s_{23} c_{13} e^{i \delta} \\[0.2cm] 
s_{12} s_{23} - c_{12} c_{23} s_{13} e^{i \delta} & 
- c_{12} s_{23} - s_{12} c_{23} s_{13} e^{i \delta} 
& c_{23} c_{13} e^{i \delta} \\ 
               \ea   \right) 
 {\rm diag}(1, e^{i \alpha}, e^{i \beta}) \, , 
\eea
where $\delta$ is a Dirac $CP$ violating phase, 
$\alpha$ and $\beta$ are possible Majorana 
$CP$ violating phases \cite{Majpha}, 
$c_{ij} = \cos\theta_{ij}$ and $s_{ij} = \sin\theta_{ij}$. 
We shall not consider the two Majorana phases in this Section. 
The angles $\theta_{12}$ and $\theta_{23}$ control the oscillations 
of solar and atmospheric neutrinos, respectively. The angle 
$\theta_{13}$ is mainly limited by reactor $\bar{\nu}_e$ experiments:
The Dirac phase $\delta$ can be measured 
in long baseline neutrino oscillation experiments 
(see, e.g., \cite{CPLBL}).\\ 

\noindent To obtain information about the PMNS matrix 
one fits \cite{SNOana1,SNOana2,carlos} the results of neutrino 
experiments to the hypothesis of neutrino oscillations. 
The relevant formula for the oscillation probabilities is 
\be \label{eq:Pab}
P(\nu_\alpha \rightarrow \nu_\beta) 
= \delta_{\alpha \beta} - 2 \, {\rm Re} \, 
\sum_{j > i} U_{\alpha i} \, U_{\alpha j}^\ast  \, 
U_{\beta i}^\ast \, U_{\beta j} \, 
\left( 1 - \exp{\frac{i \Delta m^2_{ji} \, L}{2 \, E}} \right)~,
\ee
where $\Delta m^2_{ji} = m_j^2 - m_i^2$.  
The 1 (3)$\sigma$ ranges of recent analyzes 
read in terms of the 
parametrization Eq.\ (\ref{eq:Upara}): 
\bea \label{eq:13sig}
(0.27) \, 0.35 \le \tan^2 \theta_{12} \le 0.52 \, (0.72)~ 
\cite{SNOana1}~, \\[0.3cm] 
(0.45) \, 0.75 \le \tan^2 \theta_{23} \le  1.3 \, (2.3)~, ~\cite{carlos}
\\[0.3cm] 
0 \, (0) \le \sin^2 \theta_{13} \le 0.029 \, (0.074)~ \cite{SNOana1}~.
\eea
The best--fit points are given by 
\be \label{eq:BF}
\tan^2 \theta_{12} = 0.43~ \cite{SNOana1},~ 
\tan^2 \theta_{23} = 1 ~\cite{carlos}
~. 
\ee
For $\theta_{23}$ there is an ambiguity corresponding to 
$\theta_{23} \leftrightarrow \pi/2 - \theta_{23}$, i.e., the angle lies 
on the ``light'' or ``dark side''. Matter effects in future long--baseline 
experiments will resolve this. In terms of the often used parameter 
$\sin^2 2 \theta_{23}$, which is blind to this ambiguity, 
one has at 1 (3)$\sigma$: 
$\sin^2 2 \theta_{23} \ge 0.86 \, (0.84)$. \\

\noindent Regarding the mass squared differences, 
the best--fit values are $(\dms)_{\rm BF} = 7.2 \cdot 10^{-5}$ eV$^2$ 
\cite{SNOana1}
and $(\dma)_{\rm BF} = 2.6 \cdot 10^{-3}$ eV$^2$ \cite{carlos}. 
A recent preliminary analysis of the SuperKamiokande collaboration, 
taking into account e.g.\ improved identification 
criteria and neutrino fluxes, yields a value of 
$(\dma)_{\rm BF} = 2.0 \cdot 10^{-3}$ eV$^2$ \cite{SKaachen}.  
There are two possible 
mass orderings, the normal and the inverse mass ordering: 
\bea
\baz
\mbox{ normal mass ordering:} & 
\dms = \Delta m_{21}^2 \ll 
\Delta m_{32}^2 \simeq \Delta m_{31}^2 = \dma\\[0.3cm]
\mbox{ inverse mass ordering:} & 
\dms = \Delta m_{21}^2 \ll 
-\Delta m_{31}^2 \simeq -\Delta m_{32}^2 = \dma 
\ea
\eea
Extreme cases are the normal (inverse) hierarchy with 
$m_3 = \dma \gg m_2 = \dms \gg m_1$ 
($m_2 = \dma \simeq m_1 \gg m_3 $) and the quasi--degenerate mass scheme 
for which $m_3^2 \simeq m_2^2 \simeq m_1^2 \gg \dma$. The latter is 
fulfilled for values of the neutrino masses larger than $\sim 0.2$ eV.\\ 

\noindent Ignoring the 
phases, the ``best--fit PMNS matrix'' reads:
\be
\label{eq:UBF}
U_{\rm PMNS}^{\rm BF} = 
\left(
\bad  
0.84 & 0.55 & 0 \\[0.3cm]
-0.39 & 0.59 & 0.71 \\[0.3cm]
0.39 & -0.59 & 0.71 
\ea 
\right)~.
\ee
In the pre--SNO--salt--phase analysis of 
Ref.\ \cite{carlos} there was given the 3$\sigma$ range of the 
PMNS matrix:
\be
\label{eq:U13sig}
|U_{\rm PMNS}| = 
\left(
\bad  
0.73  - 0.88 & 0.47 - 0.67 & 0 - 0.23 \\[0.3cm]
0.17 - 0.57 & 0.37 - 0.73 & 0.56 - 0.84 \\[0.3cm]
0.20 - 0.58 & 0.40 - 0.75 & 0.54 - 0.82 
\ea 
\right)~,
\ee
where the phase $\delta$ was allowed to take arbitrary values.

\noindent Next generation long--baseline experiments will be able to probe 
\dma{} and $\sin 2 \theta_{23}$ to the $\%$ accuracy \cite{LBL}. 
The element $U_{e3}$ can be probed down to the level $10^{-3}$ 
in future long--baseline or reactor experiments \cite{reac}. 
Neutrino factories \cite{nufac} can improve these bounds 
considerably. The solar neutrino mixing angle $\tan^2 \theta_{\rm sol}$ 
will see its error reduced below 10 $\%$ by experiments 
investigating the low energy neutrino fluxes from the sun \cite{low}. 

\noindent Currently no information about leptonic $CP$ violation exists. 
In oscillation experiments one can detect 
$CP$ violating effects \cite{CPLBL}, which have to be 
proportional to \cite{STP102}
\bea \label{eq:JCPlep}
J_{CP} = {\rm Im} \, \{{U_{e1} \, U_{\mu 2} \, U_{\mu 1}^\ast \, 
U_{e2}^\ast \}} = \frac{1}{8} \, \sin 2 \theta_{12} \, \sin 2 \theta_{23} 
\, \sin 2 \theta_{13} \, \cos \theta_{13} \, \sin \delta \\[0.3cm]
\simeq \frac{\D \theta_{13}}{\D 4} \left(1 - \mbox{corrections 
from $\theta_{13}$ and non--maximal $\theta_{12}$ and $\theta_{23}$} \right)~,
\eea 
where the value $\theta_{13}/4$ is the limit for small $\theta_{13}$, 
maximal $\theta_{12}$ and $\theta_{23}$.\\

\subsection{The Strategy}
As a very useful limit, the bimaximal mixing pattern \cite{bimax}, 
corresponding to $\theta_{12} = \theta_{23} = \pi/4$ and 
$\theta_{13} = 0$, can be considered. 
The resulting mixing matrix, ignoring the $CP$ violating phases, 
reads 
\be
\label{eq:Ubimax}
U_{\rm PMNS}^{\rm bimax} = 
\left(
\bad  
\frac{1}{\sqrt{2}} &  \frac{1}{\sqrt{2}} &  0 \\[0.3cm]
-\frac{1}{2} &  \frac{1}{2} &  \frac{1}{\sqrt{2}} \\[0.3cm]
\frac{1}{2} &  -\frac{1}{2} &  \frac{1}{\sqrt{2}} \\[0.3cm]
\ea 
\right)~.
\ee
The non--zero entries therefore take values of $\frac{1}{\sqrt{2}}$ and 
$\pm \frac{1}{2}$. This form of $U_{\rm PMNS}$ 
shall be our reference matrix, whose 
deviations are to be described by some small parameter $\lambda$. 
In some sense, the matrix (\ref{eq:Ubimax})
is the analogue to the unit matrix in case of quark mixing. 
Corrections of order $\lambda$ and higher to the unit matrix 
lead to the CKM matrix as parametrized in Eq.\ (\ref{eq:wolf}). 
In the same way, corrections to $U_{\rm PMNS}^{\rm bimax}$ 
of order $\lambda$ and higher can lead to the observed 
neutrino mixing phenomenology with non--maximal solar neutrino mixing, 
possible non--maximal atmospheric neutrino mixing and non--zero 
$U_{e3}$. One might state that the unit matrix in the quark sector and 
the bimaximal mixing matrix in Eq.\ (\ref{eq:Ubimax}) are the 
zeroth order form of the relevant mixing matrix. 
We shall comment on a possible origin of 
corrections to bimaximal mixing 
(along the lines of Ref.\ \cite{devbimax1}) in Section \ref{sec:consider}. 
Tries to parametrize the PMNS matrix in analogy 
to the CKM matrix suffer in general 
from the fact that from the 9 elements of $U_{\rm PMNS}$ only one is small, 
namely the element $|U_{e3}| \ls 0.27$. The other eight entries 
take typically values around 1/2 or $1/\sqrt{2}$. 
There are to our knowledge two other approaches to do something similar to the 
PMNS matrix as has been done so successfully with the CKM matrix. 
The analysis from \cite{kaus} uses 
$U_{e2} \simeq \sqrt{2} \lambda$, 
atmospheric neutrino mixing remains maximal and $U_{e3}$ is proportional to 
$\lambda^2$.   
The work \cite{zing} chooses the expansion parameter 
$\lambda = U_{\mu 3} \simeq 1/\sqrt{2}$. Solar neutrino mixing is governed 
by $U_{e2} = A \, \lambda^2$ and the small quantity $U_{e3}$ has 
to be introduced at at least the eights power of $\lambda$. Effects due to 
$CP$ violation in neutrino oscillations are --- 
courtesy of Eq.\ (\ref{eq:JCPlep}) --- 
proportional to at least 
$\lambda^{11}$. An Ansatz for \PMNS{} corresponding to 
$\tan^2 \theta_{12} = 0.5$, maximal atmospheric mixing and 
zero $U_{e3}$ has been used in \cite{zee}. To describe deviations from 
it, it has been multiplied with a Wolfenstein--like matrix.\\  

\noindent In this letter we wish to propose a purely phenomenological 
and model independent 
parametrization of the PMNS matrix by using a small 
``expansion'' parameter $\lambda$. For a useful analysis in terms of a small 
parameter one requires small quantities in the mixing matrix.  
The basic idea is given by the identification of up to three such 
small numbers in neutrino mixing phenomenology, namely: 
\begin{itemize}
\item[(i)]the deviation from maximal mixing of solar neutrinos
\item[(ii)]the small mixing element $U_{e3}$ 
\item[(iii)]the possible deviation from 
maximal mixing of atmospheric neutrinos 
\end{itemize}
Those three aspects describe all possible deviations from the bimaximal 
mixing scheme in Eq.\ (\ref{eq:Ubimax}).  
Observation (i) is now a solid experimental evidence, after 
inclusion of the SNO salt phase data \cite{SNOII}, it now holds that 
$\tan^2 \theta_{\rm \odot} < 1$ at more than 
5$\sigma$ \cite{SNOII,SNOana1,SNOana2}. 
Regarding observation (ii), only the mentioned limit of $|U_{e3}|^2 \le 0.07$ 
(at 3$\sigma$) exists. Best--fit points of three flavor analyzes of all 
neutrino data typically yield very small if not vanishing values 
for this quantity. 
Finally, atmospheric neutrino mixing insists to be 
described by solutions with a best--fit corresponding to 
maximal mixing. This remains true also when the K2K data is 
included or separately analyzed (e.g., \cite{carlos}). 
Though exactly maximal mixing and zero 
$U_{e3}$ would hint to some underlying symmetry in the lepton sector, 
one can not expect radiative corrections to allow these extreme values to  
persist down to low energy \cite{radcor}. Thus, one expects 
non--extreme values for $\theta_{13}$ and $\theta_{23}$. 
See, e.g., \cite{devbimax1,devbimax2} for ways to generate deviations from 
the bimaximal mixing scheme.

\noindent All in all, the three observations (i) to (iii) 
together with the 
mixing matrices (\ref{eq:UBF},\ref{eq:U13sig}) lead us   
to parametrize three elements of the mixing matrix as 
\bea \label{eq:master}
U_{e2} = \sqrt{\frac{1}{2}} \, (1 - \lambda)~, \\[0.3cm]
U_{e3} = A \, \lambda^n~, \\[0.3cm]
U_{\mu 3} = \sqrt{\frac{1}{2}} \, (1 - B \, \lambda^m) \, e^{i \delta} ~.
\eea
For $\lambda = 0$ 
one would have the bimaximal scheme from 
Eq.\ (\ref{eq:Ubimax})\footnote{Deviation from maximal 
solar neutrino mixing has also been analyzed in terms of the 
parameter $\epsilon = 1 - 2\, \sin^2\theta_{12}$ \cite{smir}, which 
roughly corresponds to $\lambda$: 
$\epsilon \simeq \lambda + {\cal{O}}(\lambda^2)$.}. 
The two Majorana phases are left out for the moment. 
Unitarity of \PMNS{} suffices to calculate the remaining elements.  
The parameters $A$ and $B$ are numbers of order one. 
The $\theta_{23} \leftrightarrow \pi/2 - \theta_{23}$ ambiguity 
reflects in a sign ambiguity of $B$. 
The power of $\lambda$ in the expressions for $U_{e3}$ ($U_{\mu 3}$) 
can be adjusted when more stringent limits (more precision data) 
are available.\\

\noindent We can take the best--fit value from Eq.\ (\ref{eq:BF}) to 
calculate $\lambda \simeq 0.22$, which is remarkably similar 
to the Wolfenstein parameter or the sine of the Cabibbo angle. 
The maximal allowed value of $|U_{e3}|^2 = 0.07$ corresponds to 
$U_{e3} = A \, \lambda$ with $A \simeq 1.2$. 
At 1 (3)$\sigma$, the range of $\lambda$ lies in
\be \label{eq:lamran}
\lambda \simeq (0.08) \, 0.18 - 0.28 \, (0.35)~. 
\ee
For the best--fit points of the many available analyzes \cite{SNOana2}, 
which lie in the range between 0.41 and 0.44 for $\tan^2 \theta_{12}$, 
$\lambda$ is between 0.24 and 0.22.   

\noindent If indeed $\lambda \simeq 0.22$, then 
for $m=1$ it must hold that $B \ls 0.91$ in order 
to fulfill the requirement $\sin^2 2 \theta_{23} \gs 0.85$. 
In the following we shall work with the ``best--fit'' value 
of $\lambda = 0.22$. 
If the limit on $|U_{e3}|^2$ goes below $\sim 10^{-2}$ one should 
take the power $n = 2$ in Eq.\ (\ref{eq:master}) in order to keep 
$A$ of order one. If $|U_{e3}|^2 \ls 10^{-4}$, then $n = 3$ is advantageous to 
choose. 

\noindent
Analogously, since typically (see below) 
$\sin^2 2\theta_{23} \simeq 1 - 4 \, B^2 \, \lambda^{2m} $ one 
should for values larger than $\sin^2 2\theta_{23} \simeq 0.95$ (0.99) 
use the power $m = 2$ ($m = 3$) in Eq.\ (\ref{eq:master}). Values of 
$m = 4$ would be required if a precision in $\sin^2 2\theta_{23}$ of order 
$10^{-4}$ was present, which seems improbable unless a neutrino 
factory will be operative. In terms of $\tan^2 \theta_{23}$, 
which in the future will be more appropriate to use, one will 
find that $\tan^2 \theta_{23} \simeq 1 - 4 \, B \, \lambda^m$. 
Thus, for $\tan^2 \theta_{23} \gs 0.7$ (or $\ls 1.3$) one should take 
$m=2$, while for 
$\tan^2 \theta_{23} \gs 0.9$ (or $\ls 1.1$) the value 
$m=3$ is more useful.\\

\noindent 
We shall now consider several different cases for the powers of 
$\lambda$ in Eq.\ (\ref{eq:master}). The considerations from this 
Section indicate that current data and the 
precision of future experiments on $\theta_{23}$ 
and $\theta_{13}$ limit the realistic values of $m$ and $n$ between 
1 and 3. 

\section{\label{sec:U}The Mixing Matrix}

\subsection{\label{sec:11}Case $m=n=1$}
In this case we have 
$U_{e3} = A \, \lambda$ and 
$U_{\mu 3} = 
\sqrt{\frac{1}{2}} (1 - B \, \lambda) \, e^{i \delta} $. It corresponds to 
rather large deviations from the extreme bimaximal values. 
One ``predicts'' $U_{e3}$ very close to its current limit and 
also $\sin^2 2\theta_{23}$ is on the edge of its $3 \sigma$ range. 
We identify 
\bea \label{eq:113}
\tan^2 \theta_{12} \simeq 
1 - 4 \, \lambda + 2 \, (5 + A^2) \, \lambda^2 
+ {\cal O}(\lambda^3)~,\\[0.3cm]
\bad  \sin^2 2\theta_{23} \simeq 1 - 4 \, B^2  \, \lambda^2 
+ {\cal O}(\lambda^3)~, & 
\tan^2 \theta_{23} \simeq 1 - 4 \, B \, \lambda + 2 \, (A^2 + 5 \, B^2) 
\, \lambda^2 + \ea {\cal O}(\lambda^3)\\[0.3cm]
\sin^2 \theta_{13} = A^2 \, \lambda^2~.
\eea 
The fact that $\sin^2 2\theta_{23}$ is blind to the 
$\theta_{23} \leftrightarrow \pi/2 - \theta_{23}$ ambiguity is reflected in 
the fact that $B$ appears quadratically in the last expression. 
The form of \PMNS{} is rather lengthy, and we shall give it therefore only 
to order $\lambda$:
\be \label{eq:U11}
\small 
\PMNS \simeq  \left( \bad 
\sqrt{\frac{1}{2}} \, (1 + \lambda) & 
\sqrt{\frac{1}{2}} \, (1 - \lambda) & A \, \lambda \\[0.2cm] 
-\frac{1}{2} \left(1 - (1 - B - A \, e^{i \delta}) \, \lambda \right) & 
\frac{1}{2} \left(1 + (1 + B - A \, e^{i \delta}) \, \lambda \right) &
\sqrt{\frac{1}{2}} \, (1 - B \, \lambda)\, e^{i \delta}\\[0.2cm] 
\frac{1}{2} \left(1 - (1 + B + A \, e^{i \delta}) \, \lambda \right) & 
-\frac{1}{2} \left(1 + (1 - B + A \, e^{i \delta}) \, \lambda \right) &
\sqrt{\frac{1}{2}} \, (1 + B \, \lambda)\, e^{i \delta}
\ea   \right) + {\cal O}(\lambda^2)~.
\ee
The precise form to a given order of $\lambda$ 
is easily obtained by using the unitarity of the 
mixing matrix. The corrections quadratic in $\lambda$ are 
functions of $A$ and $B$ except for $U_{e1}$, which receives only 
corrections depending on $A$. 
Important to note is that the corrections from $\lambda$ are 
responsible for the deviations from the 
values $\pm 1/2$ of the entries in the lower left 12 block. 
Finally, the invariant measure of $CP$ violation in 
neutrino oscillations is 
\be
J_{CP} = \frac{A \, \lambda}{4} 
\left( 1 - (2 + A^2 + 2 \, B^2) \, \lambda^2 
\right) \, \sin \delta + {\cal O}(\lambda^4)~. 
\ee 
Noting that $A \, \lambda \simeq \theta_{13}$,  
the corrections stemming from $\theta_{13}$ and non--maximal 
$\theta_{12, 23}$ --- as indicated in Eq.\ (\ref{eq:JCPlep}) --- 
are easily identified.  The larger the deviations from maximal 
$\theta_{12, 23}$, i.e., the larger $A$ and $B$, the smaller becomes 
$J_{CP}$. The ``prediction'' is that $CP$ violating effects are up to  
$\sim 5 \%$.  Note however that actual experiments searching for 
leptonic $CP$ violation will not just measure $J_{CP}$, see 
Section \ref{sec:LBL}.

\subsection{\label{sec:12}Case $m=1$ and $n=2$}
For these values we have 
$U_{e3} = A \, \lambda^2$ and 
$U_{\mu 3} = 
\sqrt{\frac{1}{2}} (1 - B \, \lambda) \, e^{i \delta} $. 
We can identify 
\bea \label{eq:123}
\tan^2 \theta_{12} \simeq 
1 - 4 \, \lambda + 10 \, \lambda^2 
+ {\cal O}(\lambda^3)~,\\[0.3cm]
 \bad \sin^2 2\theta_{23} \simeq 1 - 4 \, B^2  \, \lambda^2 
+ {\cal O}(\lambda^3)~, & 
\tan^2 \theta_{23} \simeq 1 - 4 \, B \, \lambda + 10 \, B^2 \, \lambda^2 + 
{\cal O}(\lambda^3)~,
\ea \\[0.3cm]
\sin^2 \theta_{13} = A^2 \, \lambda^4~.
\eea 
The ``predictions'' are $|U_{e3}|^2 \sim 10^{-3}$ and atmospheric mixing 
very close to the end of its allowed 3$\sigma$ range. 
The mixing matrix \PMNS{} reads 
\be \label{eq:U12}
\PMNS \simeq  \left( \bad 
\sqrt{\frac{1}{2}} \, (1 + \lambda) & 
\sqrt{\frac{1}{2}} \, (1 - \lambda) & A \, \lambda^2 \\[0.2cm] 
-\frac{1}{2} \left( 1 -  
(1 - B) \, \lambda  \right) & 
\frac{1}{2}  \left( 1 + 
(1 + B) \, \lambda  \right) &
\sqrt{\frac{1}{2}} \, (1 - B \, \lambda)\, e^{i \delta}\\[0.2cm] 
\frac{1}{2}  \left( 1 -  
(1 + B) \, \lambda  \right) & 
-\frac{1}{2} \left( 1 + 
(1 - B) \, \lambda  \right) &
\sqrt{\frac{1}{2}} \, (1 + B \, \lambda)\, e^{i \delta}
\ea   \right) + {\cal O}(\lambda^2)~.
\ee
It is obtained by removing the term $A \, e^{i \delta}$ from the PMNS matrix 
in the case of $m = n = 1$ as given in Eq.\ (\ref{eq:U11}). 
The corrections of order $\lambda^2$ for the 
lower left 2 by 2 submatrix are functions of $A$ and $B$. They are constant 
for $U_{e1}$ and only depending on $B$ for $U_{\tau 3}$. 
Effects of $CP$ violation are proportional to $\lambda^2$,  
\be
J_{CP} \simeq \frac{A \, \lambda^2}{4} \left( 
1 - 2 \, (1 + B^2) \, \lambda^2\right) \, \sin \delta + {\cal O}(\lambda^5)~,
\ee
and not more than a few $\%$.

\subsection{\label{sec:21}Case $m=2$ and $n=1$}
Now our parameters read 
$U_{e3} = A \, \lambda$ and 
$U_{\mu 3} = 
\sqrt{\frac{1}{2}} (1 - B \, \lambda^2) \, e^{i \delta} $. 
The mixing angles are 
\bea \label{eq:213}
\tan^2 \theta_{12} \simeq 
1 - 4 \, \lambda + 2\left( 5 + A^2 \right) \, \lambda^2 
+ {\cal O}(\lambda^3)~,\\[0.3cm]
\baz  \sin^2 2\theta_{23} \simeq 1 - (A^2  - 2 \, B)^2  \, \lambda^4 
+ {\cal O}(\lambda^5)~,& 
\tan^2 \theta_{23} \simeq 1 + 2 \, (A^2 - 2 \, B) \, \lambda^2 + 
{\cal O}(\lambda^4)~,
\ea \\[0.3cm]
\sin^2 \theta_{13} = A^2 \, \lambda^2~.
\eea  
Thus, $U_{e3}$ is close to its current limit and the deviation from 
$\sin^2 2\theta_{23} = 1$ is not more than a few $\%$. 
The mixing matrix \PMNS{} is given by  
\be \label{eq:U21}
\PMNS \simeq  \left( \bad 
\sqrt{\frac{1}{2}} \, (1 + \lambda) & 
\sqrt{\frac{1}{2}} \, (1 - \lambda) & A \, \lambda \\[0.2cm] 
-\frac{1}{2} \left( 1 -   
(1 - A \, e^{i \delta}) \, \lambda  \right) & 
\frac{1}{2}  \left( 1 +   
(1 - A \, e^{i \delta}) \, \lambda  \right) &
\sqrt{\frac{1}{2}}\, (1 - B \, \lambda^2)e^{i \delta}\\[0.2cm] 
\frac{1}{2}  \left( 1 -  
(1 + A \, e^{i \delta})\, \lambda  \right) & 
-\frac{1}{2}  \left( 1 + 
 (1 + A \, e^{i \delta}) \, \lambda  \right) &
\sqrt{\frac{1}{2}} \, (1 + B \, \lambda^2) e^{i \delta}
\ea   \right) + {\cal O}(\lambda^2) 
\ee
which is obtained from Eq.\ (\ref{eq:U11}) by removing $B$ from the lower 
left 2 by 2 submatrix. The quadratic corrections are functions of $A$ and 
$B$ except for $U_{e1}$, which only depends on $A$. 
The rephasing invariant $CP$ violation measure is  
\be
J_{CP} \simeq \frac{A \, \lambda}{4} \left( 
1 - \, (2 + A^2) \, \lambda^2 \right) \, \sin \delta 
+ {\cal O}(\lambda^4) 
\ee
being rather sizable but not exceeding $5\%$. 

\subsection{\label{sec:22}Case $m=n=2$}
Now it holds  
$U_{e3} = A \, \lambda^2$ and 
$U_{\mu 3} = 
\sqrt{\frac{1}{2}} (1 - B \, \lambda^2) \, e^{i \delta} $, 
yielding the mixing angles 
\bea \label{eq:223}
\tan^2 \theta_{12} \simeq 
1 - 4 \, \lambda + 10  \, \lambda^2 
+ {\cal O}(\lambda^3)~,\\[0.3cm]
\bad \sin^2 2\theta_{23} \simeq 1 - 4 \, B^2 \, \lambda^4 
+ {\cal O}(\lambda^5)~,& 
\tan^2 \theta_{23} \simeq 1 - 4 \, B \, \lambda^2 + {\cal O}(\lambda^3)~,
\ea \\[0.3cm]
\sin^2 \theta_{13} = A^2 \, \lambda^4~.
\eea  
The deviation from $\sin^2 2\theta_{23} = 1$ is not more than a few 
percent and $|U_{e3}|$ is on the level of $10^{-3}$. 
The mixing matrix \PMNS{} is given by  
\be \label{eq:U22}
\small
\PMNS \simeq  \left( \bad 
\sqrt{\frac{1}{2}} \, (1 + \lambda - \lambda^2) & 
\sqrt{\frac{1}{2}} \, (1 - \lambda) & A \, \lambda^2 \\[0.2cm] 
-\frac{1}{2}  \left( 1 - \, \lambda  + 
(B + A \, e^{i \delta}) \, \lambda^2 \right) & 
\frac{1}{2} \left( 1 + \lambda -   
(1 - B + A \, e^{i \delta}) \, \lambda^2  \right) &
\sqrt{\frac{1}{2}}\, (1 - B \, \lambda^2) e^{i \delta}\\[0.2cm] 
\frac{1}{2}  \left( 1 - \lambda - 
(B + A \, e^{i \delta})\, \lambda^2  \right) & 
-\frac{1}{2} \left( 1 + \lambda - 
 (1 + B - A \, e^{i \delta}) \, \lambda^2  \right) &
\sqrt{\frac{1}{2}} (1 + B \, \lambda^2) \, e^{i \delta}
\ea   \right) + {\cal O}(\lambda^3)~. 
\ee
It is seen that for the lower left 2 by 2 submatrix the linear 
corrections to the ``bimaximal'' values $\pm 1/2$ are constant 
and the quadratic ones are functions of $A$ and $B$. 
The rephasing invariant $CP$ violation measure is given by 
\be
J_{CP} \simeq \frac{A \, \lambda^2}{4} \left( 
1 - 2 \, \lambda^2 \right) \, \sin \delta + {\cal O}(\lambda^5)~,
\ee
again on the level of a few percent.

\subsection{\label{sec:23}The ``Wolfenstein Case'' $m=2$ and $n=3$}
How could one not be tempted to put the third power of the expansion 
parameter in the $U_{e3}$ and the second power in the 
$U_{\mu 3}$ element. This would resemble the Wolfenstein parametrization 
Eq.\ (\ref{eq:wolf}). In this case, i.e., 
$U_{e3} = A \, \lambda^3$ and 
$U_{\mu 3} = 
\sqrt{\frac{1}{2}} (1 - B \, \lambda^2) \, e^{i \delta} $, we have 
\bea \label{eq:233}
\tan^2 \theta_{12} \simeq 
1 - 4 \, \lambda + 10  \, \lambda^2 
+ {\cal O}(\lambda^3)~,\\[0.3cm]
\bad \sin^2 2\theta_{23} \simeq 1 - 4 \, B^2 \, \lambda^4 
+ {\cal O}(\lambda^5)~,& 
\tan^2 \theta_{23} \simeq 1 - 4 \, B \, \lambda^2 + {\cal O}(\lambda^3)~,
\ea \\[0.3cm]
\sin^2 \theta_{13} = A^2 \, \lambda^6~.
\eea 
The ``prediction'' for $\sin^2 2\theta_{23} - 1$ 
is not more than a few 
percent and $|U_{e3}|^2$ is on the level of $10^{-4}$. 
For the PMNS matrix one has 
\bea \label{eq:U23}
\PMNS \simeq  \\[0.3cm]
\small
\left( \bad 
\sqrt{\frac{1}{2}} \, (1 + \lambda - \lambda^2 + \lambda^3) & 
\sqrt{\frac{1}{2}} \, (1 - \lambda) & A \, \lambda^3 \\[0.2cm] \small
-\frac{1}{2} \, \left( 1 - \lambda  + B \, \lambda^2  
- (B - A \, e^{i \delta}) \, \lambda^3\right) & 
\frac{1}{2} \, \left( 1 + \lambda -     
(1 - B) \, \lambda^2  + (1 + B - A \, e^{i \delta}) \, \lambda^3 
\right) &
\sqrt{\frac{1}{2}}\, (1 - B \, \lambda^2) e^{i \delta}\\[0.2cm] \small
\frac{1}{2} \, \left( 1 - \lambda - B\, \lambda^2  + 
(B - A \, e^{i \delta}) \, \lambda^3 \right) &
-\frac{1}{2} \, \left( 1 + \lambda - 
 (1 + B ) \, \lambda^2  + (1 - B + A \, e^{i \delta}) \, \lambda^3\right) &
\sqrt{\frac{1}{2}} (1 + B \, \lambda^2) \, e^{i \delta}
\ea   \right) \\[0.3cm]+ {\cal O}(\lambda^4)~. 
\eea
In contrast to the Wolfenstein parametrization, however, 
$\lambda$ appears linearly in e.g.\ the $U_{\mu 2}$ element. 
Also, $CP$ violation is proportional to the third power of $\lambda$, 
\be
J_{CP} \simeq \frac{A \, \lambda^3}{4} \left( 
1 - 2 \, \lambda^2 \right) \, \sin \delta + {\cal O}(\lambda^6)~,
\ee
not exceeding one percent. 

\subsection{\label{sec:32and33} The remaining cases 
}
The remaining interesting cases are $m = 3$ with $n = 2$ and also 
$m = n = 3$. The former case --- that is 
$U_{\mu 3} = \sqrt{\frac{1}{2}}\, (1 - B \, \lambda^3) \, e^{i \delta}$  
and $U_{e3} = A \, \lambda^2$ --- yields 
\bea \label{eq:323}
\tan^2 \theta_{12} \simeq 
1 - 4 \, \lambda + 10  \, \lambda^2 
+ {\cal O}(\lambda^3)~,\\[0.3cm]
\bad  \sin^2 2\theta_{23} \simeq 1 - 4 \, B^2 \, \lambda^6 
+ {\cal O}(\lambda^7)~,& 
\tan^2 \theta_{23} \simeq 1 - 4 \, B \, \lambda^3 + {\cal O}(\lambda^4)~,
\ea \\[0.3cm]
\sin^2 \theta_{13} = A^2 \, \lambda^4~,
\eea 
with $|U_{e3}|^2 \sim 10^{-3}$ and $\sin^2 2\theta_{23} \neq 1$ 
with a precision of less than 1$\%$. 
For the PMNS matrix holds: 
 \bea \label{eq:U32}
\PMNS \simeq  \\[0.3cm] 
\hspace{-.9cm}\small
\left( \bad 
\sqrt{\frac{1}{2}} \, (1 + \lambda - \lambda^2 + \lambda^3) & 
\sqrt{\frac{1}{2}} \, (1 - \lambda) & A \, \lambda^2 \\[0.2cm] 
-\frac{1}{2} \, \left( 1 - \lambda  +  A \, e^{i \delta} \, \lambda^2  
+ (B + A \, e^{i \delta}) \, \lambda^3\right) & 
\frac{1}{2} \, \left( 1 + \lambda -     
(1 + A \, e^{i \delta}) \, \lambda^2  
+ (1 + B + A \, e^{i \delta}) \, \lambda^3 
\right) &
\sqrt{\frac{1}{2}}\, (1 - B \, \lambda^3) e^{i \delta}\\[0.2cm] 
\frac{1}{2} \, \left( 1 - \lambda - A \, e^{i \delta} \, \lambda^2  - 
(B + A \, e^{i \delta}) \, \lambda^3 \right) &
-\frac{1}{2} \, \left( 1 + \lambda - 
 (1 - A \, e^{i \delta} ) \, \lambda^2  + (1 - B - A \, e^{i \delta}) \, 
\lambda^3\right) &
\sqrt{\frac{1}{2}} (1 + B \, \lambda^3) \, e^{i \delta}
\ea   \right) \\[0.3cm]+ {\cal O}(\lambda^4)~. 
\eea
As for the case $m = n = 2$ one obtains: 
\be
J_{CP} \simeq \frac{A \, \lambda^2}{4} \left( 
1 - 2 \, \lambda^2 \right) \, \sin \delta + {\cal O}(\lambda^5) ~,
\ee
again at most a few percent.\\

\noindent  If $m = n = 3$, i.e.,  
$U_{\mu 3} = \sqrt{\frac{1}{2}} \, (1 - B \, \lambda^3)\, e^{i \delta}$  
and $U_{e3} = A \, \lambda^3$, then it holds for the mixing parameters:
\bea \label{eq:333}
\tan^2 \theta_{12} \simeq 
1 - 4 \, \lambda + 10  \, \lambda^2 
+ {\cal O}(\lambda^3)~,\\[0.3cm]
\baz \sin^2 2\theta_{23} \simeq 1 - 4 \, B^2 \, \lambda^6 
+ {\cal O}(\lambda^7)~,& 
\tan^2 \theta_{23} \simeq 1 - 4 \, B \, \lambda^3 + {\cal O}(\lambda^4)~,
\ea \\[0.3cm]
\sin^2 \theta_{13} = A^2 \, \lambda^6~,
\eea 
i.e., except for $\sin^2 \theta_{13}$ to the given order identical 
for the $m=3$ and $n=2$ case above. 
The parameters $A$ and $B$ 
appear in the mixing matrix only a third order in $\lambda$: 
\bea \label{eq:U33}
\PMNS \simeq  \\[0.3cm]
\small
\left( \bad 
\sqrt{\frac{1}{2}} \, (1 + \lambda - \lambda^2 + \lambda^3) & 
\sqrt{\frac{1}{2}} \, (1 - \lambda) & A \, \lambda^3 \\[0.2cm] 
-\frac{1}{2} \, \left( 1 - \lambda  
+ (B + A \, e^{i \delta}) \, \lambda^3 \right) & 
\frac{1}{2} \, \left( 1 + \lambda - \lambda^2  
+ (1 + B - A \, e^{i \delta}) \, \lambda^3 
\right) &
\sqrt{\frac{1}{2}}\, (1 - B \, \lambda^3) e^{i \delta}\\[0.2cm] 
\frac{1}{2} \, \left( 1 - \lambda - 
(B + A \, e^{i \delta}) \, \lambda^3 \right) &
-\frac{1}{2} \, \left( 1 + \lambda - \lambda^2  
+ (1 - B + A \, e^{i \delta}) \, 
\lambda^3\right) &
\sqrt{\frac{1}{2}} (1 + B \, \lambda^3) \, e^{i \delta}
\ea   \right) \\[0.3cm]+ {\cal O}(\lambda^4)~. 
\eea
There are no quadratic terms in the elements $U_{\mu 1}$ and $U_{\tau 1}$. 
Finally, $CP$ violation is governed by 
\be
J_{CP} \simeq \frac{A \, \lambda^3}{4} \left( 
1 - 2 \, \lambda^2 \right) \, \sin \delta + {\cal O}(\lambda^6) 
\ee
being below one percent.\\

\noindent The remaining cases $m = 3$ and $n = 1$ ($m = 1$ and $n = 3$) 
are obtained from the cases $m = 2$ and $n = 1$ ($m = 1$ and $n = 2$) 
by setting $B = 0$ ($A = 0$) in the relevant expressions for 
the mixing parameters and $J_{CP}$.

\subsection{\label{sec:consider}Speculations}
One can speculate about the origin of the corrections induced by the 
$\lambda$ terms. It is possible to imagine, e.g., that the bimaximal mixing 
scheme from Eq.\ (\ref{eq:Ubimax}) stems from the diagonalization 
of the neutrino mass matrix (this is possible, e.g., when 
a $L_e - L_\mu - L_\tau$ symmetry is present \cite{lelmlt}) 
and any corrections are implied by the 
unitary matrix $U_\ell$ that diagonalizes the charged lepton mass matrix 
\cite{devbimax1}. Recall that in a basis in which the 
charged lepton mass matrix is not diagonal the PMNS matrix 
is given by $U_\ell^\dagger \, U$, where $U$ diagonalizes the 
neutrino mass matrix in that basis. 
If we define the matrix $U_\lambda$, which induces the correction to the 
bimaximal scheme, we may write 
$U_{\rm PMNS} \equiv U_\lambda \, U^{\rm bimax}_{\rm PMNS}$, where 
$U^{\rm bimax}_{\rm PMNS}$ is given in 
Eq.\ (\ref{eq:Ubimax})\footnote{This is similar to the 
strategy in \cite{zee}, where however a different mixing matrix 
to start with was used.}. 
Then one can simply solve for $U_{\lambda}$. Taking for definiteness 
the example $m = 3$ and $n = 2$, one finds 
\bea \label{eq:Ulam}
U_{\lambda} \simeq 
\small
\left( \bad 
1 - \lambda^2/2 & 
-\lambda/\sqrt{2} + \frac{\D 1 + 2 \, A}{\D 2 \sqrt{2}} \, \lambda^2 & 
\lambda/\sqrt{2} + \frac{\D 2 \, A - 1}{\D 2 \sqrt{2}} \, \lambda^2  \\[0.2cm] 
\lambda/\sqrt{2} - \frac{\D 1 + 2 \, A \, e^{i \delta}}{\D 2 \sqrt{2}} \, 
\lambda^2 
& \frac{\D 1 + e^{i \delta}}{\D 2} - \frac{\D \lambda^2}{\D 4} &  
  \frac{\D -1 + e^{i \delta}}{\D 2} + \frac{\D \lambda^2}{\D 4} \\[0.2cm] 
-\lambda/\sqrt{2} + \frac{\D 1 - 2 \, A}{\D 2 \sqrt{2}} \, \lambda^2 & 
\frac{\D -1 + e^{i \delta}}{\D 2} + \frac{\D \lambda^2}{\D 4} & 
\frac{\D 1 + e^{i \delta}}{\D 2} - \frac{\D \lambda^2}{\D 4}\\[0.2cm] 
\ea   \right) + {\cal O}(\lambda^3)~. 
\eea
It is seen that 23 and 32 entries are in general of order one but reduce to 
order $\lambda^2$ for $CP$ conservation. 
Those entries can also be of order $\lambda$, however only for for the 
cases $m = n = 1$ and $m=1$ with $n=2$.   
Thus, if $CP$ is conserved and atmospheric neutrino mixing 
is very close to maximal, the 
matrix $U_\lambda$ takes the unit matrix as the dominant form with 
corrections of order $\lambda$. The typical ``CKM--structure'' with 
very small $\lambda^3$ terms is however not necessary.

\section{\label{sec:mnu}The Majorana Mass matrix}

\subsection{\label{sec:mnubas}Basics}
Up to know our analysis assumed only the neutrino oscillation 
explanation of the experimental data. Now we assume in addition that 
neutrinos are Majorana particles, which 
is, e.g., a prediction of the see--saw mechanism \cite{seesaw}. 
Thus, the neutrino mass matrix $m_\nu$ 
in the basis in which the charged lepton mass matrix is diagonal 
is given by: 
\begin{equation} \label{eq:mnu}
m_\nu = U_{\rm PMNS} \, \, m_\nu^{\rm diag} \, \, U_{\rm PMNS}^T~.
\end{equation}
Here $m_\nu^{\rm diag}$ is a diagonal 
matrix containing the masses $m_{1,2,3}$
of the three massive Majorana neutrinos. Immediate consequence of the 
Majorana nature of the neutrinos is the presence of two Majorana 
phases $\alpha$ and $\beta$ to which neutrino oscillations 
are insensitive \cite{Majins}. 
Information about these phases can be obtained
by studying processes in which the total lepton charge $L$ 
changes by two units, e.g., 
\onbb, $K^{+} \rightarrow \pi^{-} + \mu^{+} + \mu^{+}$, 
etc. Realistically, only \onbb{} can expected to be measured 
\cite{ichL}. The decay width of this process is sensitive 
to the $ee$ element of $m_\nu$.\\ 

\noindent An interesting observation is that the ratio of typical best--fit 
values of the mass squared differences corresponds roughly to the 
expansion parameter $\lambda$:
\be \label{eq:obsdm}
R \equiv \sqrt{\frac{(\dms)_{\rm BF}}{(\dma)_{\rm BF}}} \simeq 
\sqrt{\frac{7.2 \cdot 10^{-5}}{2.0 \cdot 10^{-3}}} \simeq 0.19 \sim 
\lambda ~.
\ee 
We took for \dma{} the best--fit point of the  
preliminary new analysis of the SuperKamiokande collaboration 
\cite{SKaachen}. 
Using for \dma{} the 
90 $\%$ C.L.\ analysis from \cite{SKaachen}, which is  
$(1.3 \ldots 3.1) \cdot 10^{-3}$ 
eV$^2$, with the 90 $\%$ C.L.\ range of \dms{} from \cite{SNOana1}, we 
find that $R$ lies between 0.13 and 0.28. This corresponds to 
a good precision to the 3$\sigma$ range of $\lambda$ as given in 
Eq.\ (\ref{eq:lamran}). 
In the following we shall 
assume that $R \simeq \lambda$ and study the resulting structure of the 
neutrino mass matrix. The results do not change much unless 
\dma, \dms{} and $\tan^2 \theta_{12}$ are on the very edges of their allowed 
ranges. 
Before we perform this analysis, 
it is useful to study the mass matrix again in the limit 
of exact bimaximal mixing. 
In the following, we will neglect the $CP$ violating phases, see, e.g.,  
\cite{frig} for an analysis of the structure of $m_\nu$ in case of 
complex entries.   
Using Eqs.\ (\ref{eq:Upara}) and (\ref{eq:mnu}), the mass matrix reads 
\bea \label{eq:mnubimax}
m_\nu = \left( \bad 
\D A & B & - B \\[0.2cm] 
\D \cdot & \D D + \frac{A}{2} &  \D D - \frac{A}{2} \\[0.3cm]
\D \cdot & \cdot & \D D + \frac{A}{2} 
\ea   \right)~, 
\eea
where 
\be
A = \frac{m_1 + m_2}{2}~,~
B = \frac{m_2 - m_1}{2 \, \sqrt{2}}~
,~D = \frac{m_3}{2}~. 
\ee
As mentioned, there are three 
extreme cases for the mass hierarchies, the 
normal hierarchy (NH) with $\sqrt{\dma} = m_3 \gg m_2 \simeq \sqrt{\dms} 
\gg m_1 \simeq 0$, 
the inverse hierarchy (IH) with $\sqrt{\dma} = 
m_2 \simeq m_1 \gg m_3 \simeq 0$ 
and quasi--degenerate neutrinos (QD) with $m_0 \equiv 
m_3 \simeq m_2 \simeq m_1$. 
Depending on the relative signs of the mass states, several 
extreme forms of the mass matrix result. 
In case of NH, one finds for $m_{1,2} = 0$:  
\bea \label{eq:NHex}
m_\nu = \D \frac{\sqrt{\dma}}{2} \, \left( \bad 
\D 0 & 0 & 0 \\[0.2cm] 
\D \cdot & \D 1  &  \D 1 \\[0.3cm]
\D \cdot & \cdot & \D 1 
\ea   \right)~, 
\eea
i.e., a mass matrix with a leading $\mu\tau$ block. 
Regarding IH, the third mass $m_3$ can safely be 
neglected. The form of $m_\nu$ then depends on the relative sign of the 
two mass states $m_1$ and $m_2$: 
\bea \label{eq:IHex}
m_\nu = \sqrt{\dma} \,  
\left\{ 
\baz  
\left( 
\bad 
1 & 0 & 0 \\[0.3cm]
\cdot & \frac{1}{2} & -\frac{1}{2} \\[0.3cm]
\cdot & \cdot & \frac{1}{2} \ea \right) & \mbox{ same sign}  \\[0.5cm]
\left( 
\bad 
0 & \frac{1}{\sqrt{2}} & -\frac{1}{\sqrt{2}} \\[0.3cm]
\cdot &  0 &  0 \\[0.3cm]
\cdot & \cdot & 0 \ea \right) &  \mbox{opposite sign}
\ea
\right.
\eea
For the QD spectrum one finds 
\bea \label{eq:QDex}
m_\nu = m_0 \,  
\left\{ 
\baz  
\left( 
\bad 
1 & 0 & 0 \\[0.3cm]
\cdot & 1 & 0 \\[0.3cm]
\cdot & \cdot & 1 \ea \right) & 
{\rm sign}(m_1) = {\rm sign}(m_2) = {\rm sign}(m_3)\\[0.5cm]
\left( 
\bad 
0 & -\frac{1}{\sqrt{2}} & \frac{1}{\sqrt{2}} \\[0.3cm]
\cdot &  \frac{1}{2} &  \frac{1}{2} \\[0.3cm]
\cdot & \cdot & \frac{1}{2} \ea \right) &  
{\rm sign}(m_1) = -{\rm sign}(m_2) = {\rm sign}(m_3)\\[0.5cm]
\left( 
\bad 
0 & \frac{1}{\sqrt{2}} & -\frac{1}{\sqrt{2}} \\[0.3cm]
\cdot &  \frac{1}{2} &  \frac{1}{2} \\[0.3cm]
\cdot & \cdot & \frac{1}{2} \ea \right) &  
{\rm sign}(m_1) = -{\rm sign}(m_2) = -{\rm sign}(m_3)\\[0.5cm]
\left( 
\bad 
1 & 0 & 0 \\[0.3cm]
\cdot & 0 & -1 \\[0.3cm]
\cdot & \cdot & 0 \ea \right) & 
{\rm sign}(m_1) = {\rm sign}(m_2) = -{\rm sign}(m_3)  \\[0.5cm]
\ea
\right.
\eea
We can expect that in our parametrization the parameter 
$\lambda$ will appear in the neutrino mass matrix 
at least linearly in order to correct the extreme 
values $0, \pm \frac{1}{\sqrt{2}}, \pm 1$ and $\pm 1/2$.

\subsection{\label{sec:NH}Normal hierarchy}
In case of the normal hierarchy we have 
\be \label{eq:massNH}
m_3 = \sqrt{\dma}~,~m_2 = \sqrt{\dms} \, \lambda\mbox{ and } 
m_1 = D \, \sqrt{\dma} \, \lambda^{2+l}~,\mbox{ where } l \ge 0~.
\ee
The expression for $m_1$ with $D = {\cal{O}}(1)$ expresses our 
lacking knowledge of it. 
A similar Ansatz for the 
structure of $m_\nu$ in case of a normal hierarchical mass scheme 
has been made in \cite{kaus}. 
For $m = n = l = 1$ and all mass states positive the mass matrix looks like 
\be \small \label{eq:mnuNH}
m_\nu = \frac{\sqrt{\dma}}{2}
\left( 
\bad 
\lambda 
& \frac{1 + 2 \, A}{\sqrt{2}} \, \lambda  
& \frac{2 A - 1}{\sqrt{2}} \, \lambda  
\\[0.3cm]
\cdot & 1 + \left(\frac{1}{2} - 2 \, B \right) \, \lambda  
& 1 - \frac{\lambda}{2} 
\\[0.3cm]
\cdot & \cdot & 1 + \left( \frac{1}{2} + 2 \, B \right) \, \lambda 
\ea
\right) ~,
\ee
neglecting terms of order ${\cal O}(\lambda^2)$. 
The characteristic ``leading $\mu\tau$ block'' structure of 
$m_\nu$ from Eq.\ (\ref{eq:NHex}) is identified. 
Corrections at order $\lambda$ depend on $A$ in the $e$--row 
of $m_\nu$ and on $B$ for the $\mu\mu$ and $\tau\tau$ entries. 
Higher powers of $\lambda$ in $m_1$ will --- to order $\lambda^2$ --- 
lead to the disappearance of $D$ in the formula for $m_\nu$. 
Setting $U_{e3} = A \, \lambda^2$ leads to a mass matrix in which 
to order $\lambda^2$ the parameter $A$ does not appear in the $\mu\tau$ 
submatrix as well as in the $ee$ entry. It is obtained by removing $A$ 
from the last equation in the indicated entries. 
The matrix for $U_{\mu 3} = \sqrt{\frac{1}{2}} \, (1 - B \, \lambda^2)$ 
but $U_{e3} = A \, \lambda$ is given by Eq.\ (\ref{eq:mnuNH}) by 
removing $B$ from the first row and from the linear terms of the $\mu\mu$  
and $\tau\tau$ entries. 
The ``Wolfenstein like'' parametrization with $U_{e3} = A \, \lambda^3$ 
and $U_{\mu 3} = \sqrt{\frac{1}{2}} \, (1 - B \, \lambda^2)$ together 
with $m_1 = D \, \sqrt{\dma} \, \lambda^3$ leads to the particularly 
simple form 
\be  \label{eq:mnuNH1}
m_\nu = \frac{\sqrt{\dma}}{2}
\left( 
\bad 
\lambda - 2 \, \lambda^2 & \frac{\lambda}{\sqrt{2}} & 
- \frac{\lambda}{\sqrt{2}} \\[0.3cm]
\cdot & 1 + \frac{\lambda}{2} + (1 - 2 \, B) \, \lambda^2 & 
1 - \frac{\lambda}{2} - \lambda^2 \\[0.3cm]
\cdot & \cdot & 1 + \frac{\lambda}{2} + (1 + 2 \, B) \, \lambda^2 
\ea
\right) +  
{\cal O}(\lambda^3)~.
\ee
The  $\theta_{23} \leftrightarrow \pi/2 - \theta_{23}$ ambiguity, 
which translates into a sign ambiguity of $B$, is seen to have 
origin in the size of the $\mu\mu$ and $\tau\tau$ entries. E.g., 
for all masses positive and 
$\theta_{23} > \pi/4$ the $\tau\tau$ entry is larger. 
The structure of the mass matrix 
does not depend on the exactness of the relation $R = \lambda$ or the 
relative signs of the mass states. When 
${\rm sign} (m_1) = - {\rm sign} (m_2) = - {\rm sign} (m_3)$, then 
the mass matrix looks as above. 
For ${\rm sign} (m_1) = - {\rm sign} (m_2) = {\rm sign} (m_3)$ 
and ${\rm sign} (m_1) = {\rm sign} (m_2) = {-\rm sign} (m_3)$ one has 
to replace\footnote{The convention here and in the 
following will be such that the sign of the $ee$ entry is positive.} 
$A$ with $-A$, $B$ with $-B$ and the 1 in the $\mu\tau$ block with $-1$.    
If we further 
choose $U_{\mu 3} = \sqrt{\frac{1}{2}} \, (1 - B \, \lambda^3)$ than we 
obtain a mass matrix in which up to order $\lambda^2$ the parameters 
$A, B$ and $D$ do not appear at all. It is obtained by setting in the 
last equation $B=0$.

\subsection{\label{sec:IH}Inverse hierarchy}
In this case we have 
\be \label{eq:massIH}
m_2 = \sqrt{\dma}~,~m_1 \simeq  \sqrt{\dma} \, (1 - \lambda^2/2) 
\mbox{ and } m_3 = D \, \sqrt{\dma} \, \lambda^{2+l}~,\mbox{ where } l \ge 0~.
\ee
The dependence on the power of $\lambda$ in $m_3$ 
is almost vanishing. 
The form of $m_\nu$ depends strongly on the signs of the masses $m_1$ and 
$m_2$.  
For identical 
relative signs between $m_1$ and $m_2$, the $e\mu$ and $e \tau$ entries 
are suppressed by $\lambda$ or $\lambda^2$, depending on the powers 
of $\lambda$ in $U_{\mu 3}$ or $U_{e3}$. 
If e.g., $m = 2$ and $n = 1$ or $m = n = 1$, then the entries are 
of order $\lambda$. For all other cases under consideration, 
these terms are of order $\lambda^2$. 
The remaining independent entries of $m_\nu$ are order one.
If $m_1$ and $m_2$ have opposite relative signs, then $m_{e \mu}$ 
and $m_{e \tau}$ are order one and the remaining entries of $m_\nu$ are 
linear in $\lambda$, independent of $m$ and $n$. 
One finds for $m = n = 1$ that for same signs of $m_1$ and $m_2$ 
\be 
m_\nu \simeq \sqrt{\dma} \, 
\left( 
\bad 
1  
& - \frac{A}{\sqrt{2}} \lambda 
& - \frac{A}{\sqrt{2}} \lambda 
\\[0.3cm]
\cdot & \frac{1}{2} + B \, \lambda 
& -\frac{1}{2} + B \, \lambda   
\\[0.3cm]
\cdot & \cdot & \frac{1}{2} - B \, \lambda  
\ea \right) + {\cal{O}}(\lambda^2)~, 
\ee
while for opposite signs
\be 
m_\nu \simeq \sqrt{\dma} \, 
\left( 
\bad 
2 \, \lambda  & 
-\frac{1}{\sqrt{2}} - \frac{B}{\sqrt{2}} \, \lambda  
& \frac{1}{\sqrt{2}} - \frac{B}{\sqrt{2}} \, \lambda  
\\[0.3cm]
\cdot &  (A - 1) \, \lambda 
&  \lambda  
\\[0.3cm]
\cdot & \cdot & -(A + 1) \, \lambda 
\ea \right) + {\cal{O}}(\lambda^2)~.
\ee
The parameter $\lambda$ appears at least linearly 
to correct the extreme ``bimaximal'' mass matrices from Eq.\ (\ref{eq:IHex}). 
For same signs, $A$ appears in the $e$--row and $B$ in the $\mu \tau$ sector, 
whereas for opposite signs it is vice versa. 
Taking as another example again the ``Wolfenstein--like'' case 
$m = 2$ and $n = 3$ one finds to order $\lambda^2$: 
\bea \small 
m_\nu \simeq \sqrt{\dma} \, 
\left\{ \baz 
\left( 
\bad
1 - \frac{1}{4} \, \lambda^2 
& \frac{1}{4 \sqrt{2}} \, \lambda^2 
& -  \frac{1}{4 \sqrt{2}} \, \lambda^2 \\[0.3cm]
\cdot & \frac{1}{2} + (B - \frac{1}{8}) \, \lambda^2  
& -\frac{1}{2} +  \frac{1}{8} \, \lambda^2   \\[0.3cm]
\cdot & \cdot & \frac{1}{2} -  (B + \frac{1}{8}) \, \lambda^2  
\ea \right) & \mbox{same signs}\\[0.5cm]
\left( 
\bad 
2 \, \lambda - \frac{5}{4} \, \lambda^2 & 
-\frac{1}{\sqrt{2}} + \frac{9 - 4\, B}{4\sqrt{2}} \, \lambda^2  
&  \frac{1}{\sqrt{2}} - \frac{9 + 4\, B}{4\sqrt{2}} \, \lambda^2  \\[0.3cm]
\cdot &  -\lambda + \frac{3}{8} \, \lambda^2 
&  \lambda - \frac{3}{8} \, \lambda^2  \\[0.3cm]
\cdot & \cdot & - \lambda + \frac{3}{8} \, \lambda^2 
\ea \right) & \mbox{opposite signs} 
\ea \right. 
\eea
As usual, $B$ will not show up for higher orders of $\lambda$ in $U_{\mu 3}$, 
leading to a mass matrix that is at order $\lambda^2$ 
only a function of $\lambda$. It is again obtained by setting in the 
last equation $B=0$.

\subsection{\label{sec:QD}Quasi--degenerate Neutrinos}
For quasi--degenerate neutrinos, i.e., $m_3^2 \simeq m_2^2 
\simeq m_1^2 \gg \dma$, there is another small quantity introduced, namely 
the ratio of the common mass scale $m_0$ with \dma. 
For simplicity we work with the normal mass ordering. In this case we 
can express the three mass states as  
\bea \label{eq:massQD}
m_3 \equiv m_0 ~,~m_2 = a \, m_0 \mbox{ and } m_1 = b \, m_0~, \\[0.3cm]
\mbox{ where } a = 1 - \eta ~,~ b = 1 - \eta \, (1 + \lambda^2) \\[0.3cm]
\mbox{ and } \eta = \frac{\dma}{\D 2 \, m_0^2}~.
\eea
The common mass scale is denoted by $m_0$. 
These expressions for the masses are valid to order $\eta$. 
Since the spectrum is quasi--degenerate for 
$m_0 \gs 0.2$ eV, we can estimate $\eta \ls 0.04$ eV, therefore 
$\lambda^2 > \eta$. 

\noindent First, we take the case that all mass states have the 
same relative sign. For $m = n = 1$ we find: 
\bea 
(m_\nu)_{+++} = m_0 \,  
\left(  \bad 
1 - \eta & \frac{A}{\sqrt{2}} \, \eta \, \lambda & 
\frac{A}{\sqrt{2}} \, \eta \, \lambda \\[0.3cm]
\cdot & 1 - \frac{\eta}{2} - B \, \eta \, \lambda & \frac{\eta}{2}\\[0.3cm]
\cdot & \cdot & 1 - \frac{\eta}{2} + B \, \eta \, \lambda 
\ea 
\right) 
+ {\cal O}(\eta \, \lambda^2) ~.
\eea
Taking the case 
$m=2$ and $n=3$ we have
\bea 
(m_\nu)_{+++} = m_0 \,  
\left(  \bad 
1 - \eta - \frac{\eta \, \lambda^2}{2} & 
\frac{\eta \, \lambda^2}{2 \sqrt{2}} & 
-\frac{\eta \, \lambda^2}{2 \sqrt{2}} \\[0.3cm]
\cdot & 1 - \frac{\eta}{2} - \frac{1 + 4 \, B}{4} \, \eta \, \lambda^2 & 
\frac{\eta}{2} + \frac{\eta \, \lambda^2}{4}  \\[0.3cm]
\cdot & \cdot & 1 - \frac{\eta}{2} + \frac{4 \, B - 1}{4} 
\, \eta \, \lambda^2 
\ea 
\right) 
+ {\cal O}(\eta \, \lambda^3) ~.
\eea
It is seen that for $\eta=0$ the mass matrix is proportional to the 
unit matrix, irrespective of $m$ and $n$. The corrections to the zero 
values of the extreme bimaximal form from Eq.\ (\ref{eq:QDex}) 
are very small.\\

\noindent If 
${\rm sign}(m_1) = -{\rm sign}(m_2) = {\rm sign}(m_3)$, then 
the dependence on $\eta$ is not so important. 
Neglecting $\eta$ with respect to terms of order 1, we find for 
$m = n = 1$:
\bea 
(m_\nu)_{+-+} = m_0 \,  
\left(  \bad 
2 \lambda & - \frac{1}{\sqrt{2}} + \frac{A - B}{\sqrt{2}} \, \lambda & 
\frac{1}{\sqrt{2}} +  \frac{A - B}{\sqrt{2}} \, \lambda \\[0.3cm] 
\cdot & \frac{1}{2} + (A - B - 1) \, \lambda & 
\frac{1}{2} + \lambda \\[0.3cm]
\cdot & \cdot & \frac{1}{2} -  (A + 1 - B) \, \lambda 
\ea 
\right) 
+ {\cal O}(\eta, \lambda^2) 
\eea
and for $m = 2$, $n=3$:
\bea 
(m_\nu)_{+-+} = m_0 \,  
\left(  \bad 
2 \lambda - \lambda^2 & 
- \frac{1 - \eta}{\sqrt{2}} + \frac{2 - B}{\sqrt{2}} \, \lambda^2 & 
\frac{1-\eta}{\sqrt{2}} -  \frac{B + 2}{\sqrt{2}} \, \lambda^2 \\[0.3cm] 
\cdot & \frac{1}{2} - \lambda + \left(\frac{1}{2} - B \right) \, \lambda^2 
& \frac{1}{2} + \lambda - \frac{1}{2} \, \lambda^2 \\[0.3cm]
\cdot & \cdot & \frac{1}{2} -  \lambda + 
\left(\frac{1}{2} + B\right) \, \lambda^2
\ea 
\right) 
+ {\cal O}(\eta\,\lambda,  \lambda^3) ~.
\eea
For the case 
${\rm sign}(m_1) = -{\rm sign}(m_2) = -{\rm sign}(m_3)$ one finds 
for $m = n = 1$ 
\bea 
(m_\nu)_{+--} = m_0 \,  
\left(  \bad 
2 \lambda & - \frac{1}{\sqrt{2}} - \frac{A + B}{\sqrt{2}} \, \lambda & 
\frac{1}{\sqrt{2}} -  \frac{A + B}{\sqrt{2}} \, \lambda \\[0.3cm] 
\cdot & -\frac{1}{2} + (A + B - 1) \, \lambda & 
-\frac{1}{2} + \lambda \\[0.3cm]
\cdot & \cdot & -\frac{1}{2} -  (A + B + 1) \, \lambda 
\ea 
\right) 
+ {\cal O}(\eta, \lambda^2) 
\eea
and for $m = 2$, $n=3$:
\bea 
(m_\nu)_{+--} = m_0 \,  
\left(  \bad 
2 \lambda - \lambda^2 & 
- \frac{1 - \eta}{\sqrt{2}} + \frac{2 - B}{\sqrt{2}} \, \lambda^2 & 
\frac{1 - \eta}{\sqrt{2}} -  \frac{B + 2}{\sqrt{2}} \, \lambda^2 \\[0.3cm] 
\cdot & -\frac{1}{2} - \lambda + \left(\frac{1}{2} + B\right) \, \lambda^2 & 
-\frac{1}{2} + \lambda - \frac{1}{2} \, \lambda^2 \\[0.3cm]
\cdot & \cdot & -\frac{1}{2} - \lambda + \left(\frac{1}{2} - B\right) 
\, \lambda^2
\ea 
\right) 
+ {\cal O}(\eta \, \lambda, \lambda^3) 
\eea
These two last cases look very similar. 
Finally, the situation for 
${\rm sign}(m_1) = {\rm sign}(m_2) = -{\rm sign}(m_3)$ looks simpler: 
e.g., when $m = n = 1$: 
\bea 
(m_\nu)_{++-} = m_0 \,  
\left(  \bad 
1 - 2 \, A^2 \,  \lambda^2 & 
- \sqrt{2} \, A \, \lambda + \sqrt{2} \, A \, B \, \lambda^2 & 
- \sqrt{2} \, A \, \lambda - \sqrt{2} \, A \, B \, \lambda^2 \\[0.3cm] 
\cdot & 2 \, B \, \lambda - B^2 \, \lambda^2 & 
-1 + (A^2 + 2 \, B^2 )  \, \lambda^2 \\[0.3cm]
\cdot & \cdot & - 2 \, B \, \lambda + (2 \, A^2 + B^2) \, \lambda^2
\ea 
\right) 
+ {\cal O}(\eta, \lambda^3) ~.
\eea
The corrections to the entry 1 (0) that is present in the extreme 
bimaximal form from Eq.\ (\ref{eq:QDex}) are at least quadratic 
(linear).  
For $m = 2$ and $n=3$ it holds: 
\bea 
(m_\nu)_{++-} = m_0 \,  
\left(  \bad 
1 - \eta - \frac{1}{2} \, \eta \,  \lambda^2 & 
\frac{1}{2\sqrt{2}} \, \eta \,  \lambda^2 & 
- \frac{1}{2\sqrt{2}} \, \eta \,  \lambda^2 \\[0.3cm] 
\cdot & -\frac{\eta}{2} + \left(B \, (2 - \eta) - 
\frac{\eta}{4} \right) \, \lambda^2 &  
-1 + \frac{\eta}{2} + \frac{1}{4} \, \eta \,  \lambda^2\\[0.3cm] 
\cdot & \cdot & - \frac{\eta}{2} - \left(B \, (2 - \eta) + \frac{\eta}{4} 
 \right) \, \lambda^2
\ea 
\right) 
+ {\cal O}(\lambda^3) 
\eea

\subsection{\label{sec:mnusum}Summary for the Mass Matrix}
\noindent Looking at the cases considered in the last Subsections, the 
following summarizing statements can be made: 
\begin{itemize}
\item Roughly, for $|U_{e3}| \sim 0.01$ and 
$\sin^2 2\theta_{23} \ls 0.9$, 
corrections to the extreme forms of the mass matrices 
in Eqs.\ (\ref{eq:NHex},\ref{eq:IHex},\ref{eq:QDex}) are linear in 
$\lambda$. 
When $|U_{e3}| \ls 10^{-3}$ and $\sin^2 2\theta_{23} \gs 0.95$, the 
corrections become quadratic. 
\item For the normal hierarchy, corrections to the exact bimaximal form 
are at least order $\lambda$. To lowest order, the 
parameter $A$ appears in the $e$ row of $m_\nu$ and $B$ in the 
$\mu\mu$ and $\tau\tau$ elements. 
The number $D$, parameterizing the unknown 
lightest mass state, appears in all entries. There is basically no 
dependence on the relative signs of the mass states. 
\item For the inverse hierarchy, the dependence on $D$ is 
highly suppressed. For identical signs of the two heaviest mass states, 
the correction to the $ee$ entry, whose 
extreme value in case of bimaximal mixing is (in units of \dma) 1, 
is at least order $\lambda^2$. The remaining elements receive 
at least linear corrections. 
$A$ appears at leading order 
in the $e$--row and $B$ in the $\mu \tau$ sector. 
Opposite signs of the two leading mass states lead to linear 
corrections to the entries and the appearance of $A$ in the 
$\mu \tau$ sector and $B$ in the $e$--row. 
\item In case of a quasi--degenerate spectrum and identical signs of the 
masses, the corrections to the unit matrix are at least quadratical. 
The cases ${\rm sign}(m_1) = -{\rm sign}(m_2) = {\rm sign}(m_3)$ 
and ${\rm sign}(m_1) = -{\rm sign}(m_2) = -{\rm sign}(m_3)$ look 
very similar. For ${\rm sign}(m_1) = {\rm sign}(m_2) = -{\rm sign}(m_3)$ 
there are only quadratic corrections to the $ee$ and $\mu \tau$ entries. 
\item The $\mu\tau$ entry is special in our parametrization since the 
parameters $A$ and $B$ do typically only appear there for rather 
large deviations from zero $U_{e3}$ and from maximal atmospheric neutrino 
mixing.   
\item If we would consider the inverse ordering in the QD mass spectrum, 
one has to change Eq.\ (\ref{eq:massQD}) to 
$m_2 = m_0$, $m_1 = a \, m_0$ and $m_3 = b \, m_0$. 
Only simple sign changes for $A$ and/or $B$ in some elements 
of $m_\nu$ would be the result. The main difference would be for the case 
${\rm sign}(m_1) = {\rm sign}(m_2) = {\rm sign}(m_3)$, where 
the corrections to the $e\mu$ and $e\tau$ elements are now order 
$\eta$ and not just $\eta \,\lambda$ or $\eta \,\lambda^2$.  
\end{itemize}

\section{\label{sec:app}Applications}
It is surely useful to study formulae which are obtained by expansions in 
small parameters or by certain simplifications within our parametrization. 
We shall perform this analysis now for the effective mass as 
measurable in  \onbb{} and the oscillation probabilities for 
long--baseline neutrino oscillations. 

\subsection{\label{sec:0vbb}Neutrinoless Double Beta Decay}
We shall analyze now within our parametrization 
the form of the $ee$ element of $m_\nu$, which is denoted by \meff. 
In a given mass scheme or hierarchy one can considerably simplify 
the expression for \meff{} \cite{0vbb}. 
We first note that since 
\be \label{eq:meff}
\meff = \left| \sum m_i \, U_{ei}^2 \right|~,
\ee
the results are independent on the power of $\lambda$ in $U_{\mu 3}$. 

\subsubsection{\label{sec:0vbbNH}Normal Hierarchy}
With the help of Eqs.\ (\ref{eq:massNH},\ref{eq:meff}) we can 
evaluate the effective mass in case of the normal hierarchy. 
We find for $U_{e3} = A \, \lambda$ and $m_3 = D \, \lambda^2$:
\be \label{eq:meffNH1}
\meff \simeq \frac{\sqrt{\dma}}{2} \, 
\left[ \lambda - 2 \, \lambda^2 \, 
\left( 
1 - \frac{D}{2} \, c_{2 \alpha} - A^2 \, c_{2 (\alpha - \beta)} 
\right)  \right]+ {\cal O}(\lambda^3)~, 
\ee 
where $c_{2 \alpha} = \cos 2\alpha$ and so on. Terms of order 
$\lambda^3$ depend on $A, D$ and the two Majorana phases. 
Choosing $m_3 = D \, \lambda^3$ or higher powers of $\lambda$ leads to 
the disappearance of $D$ in the formula. 

\noindent 
For $m_3 = D \, \lambda^2$ and higher orders of $\lambda$ in $U_{e3}$, 
i.e., $n \ge 2$, it holds 
\be \label{eq:meffNH2}
\meff \simeq \frac{\sqrt{\dma}}{2} \, \left[ 
\lambda - 2 \, \lambda^2 \, 
\left( 
1 - \frac{D}{2} \, c_{2 \alpha}\right) + \frac{\lambda^3}{4} \, 
\left( 
4 \, (1 + 2 \, D \, c_{2 \alpha}) + D^2 \, (1 - c_{4 \alpha}) \right)
\right]
+ {\cal O}(\lambda^4) ~.
\ee 
The formulae for $m_3 = D \, \lambda^3$ correspond to setting 
$D^2 = 0$ in this last equation. 
Roughly, we can estimate the effective mass in the normal 
hierarchy as 
\be \label{eq:scaleNH}
\meff \sim \frac{\lambda}{2} \, \sqrt{\dma} \ls 0.005 \, \rm eV~.
\ee

\subsubsection{\label{sec:0vbbIH}Inverse Hierarchy}
From Eqs.\ (\ref{eq:massIH},\ref{eq:meff}) one sees that in the 
expression for \meff{} the dependence on $\beta$ practically vanishes. 
The result for $U_{e3} = A \, \lambda$ is 
\be \label{eq:meffIH}
\meff \simeq \sqrt{\dma} \sqrt{c_\alpha^2 + \frac{1}{4} 
\left(7 - 4 \, A^2 - (9 + 4 \, A^2) c_{2 \alpha} \right) \, \lambda^2} 
+ {\cal O}(\lambda^3)~.
\ee
Higher powers of $\lambda$ in $U_{e3}$ lead to the disappearance of $A$ 
in this equation. The maximal and minimal values are obtained 
when $\alpha$ takes the values 0 and $\pi/2$, respectively. Thus, 
\be \label{eq:rangeIH}
  2 \, \sqrt{\dma} \, \lambda  \ls \meff \ls 
\sqrt{\dma} \left(1 - \left(A^2 + \frac{1}{4} \right)\lambda^2 \right)~,  
\ee
up to corrections of ${\cal O}(\lambda^3)$. For no extreme values of 
$\alpha$, the scale of the effective mass is 
\be \label{eq:scaleIH}
\meff \sim \sqrt{\dma} \gs 0.05~ \rm eV~.
\ee
Comparing this with the value of \meff{} in 
the normal hierarchy in Eq.\ (\ref{eq:scaleNH}) one sees that 
the expansion parameter $\lambda$ shows up as the 
ratio of the typical values 
of \meff{} in the inverted and normal hierarchy.

\noindent It is known  
that extraction of information from a measurement of \obb{} suffers 
from a large uncertainty stemming from the calculation 
of the nuclear matrix elements. This uncertainty is 
a number of order one \cite{NME}.  
It is therefore an important question to ask and an even more important one 
to answer whether 
future \obb{} experiments can distinguish \cite{0vbb,carlos2} 
between the normal and 
inverted mass hierarchy. Let us parametrize the nuclear matrix element 
uncertainty with a factor $\xi$ as done in \cite{PPR}. 
In order to distinguish the normal from the inverted hierarchy it must hold 
that the maximal value of \meff{} in the normal hierarchy times the 
uncertainty $\xi$ has to be smaller than the minimal value of \meff{} in 
the inverted hierarchy. Therefore, choosing $U_{e3} = A \, \lambda$ and 
small $m_3$ we find from Eqs.\ (\ref{eq:meffNH1},\ref{eq:rangeIH})
\be \label{eq:xiNHIH}
\xi \ls 4 \, \left(1 + 2 \, \lambda (1 \pm A^2) \right) 
+ {\cal O}(\lambda^3)~.
\ee
Needless to say, $A$ vanishes for smaller values of $U_{e3}$. 
If that is the case, then $\xi \ls 6$, which is a very realistic number. 
Thus, with our expansion parameter $\lambda \simeq 0.2$ 
and $|U_{e3}|^2 \ls 0.01$ 
it is easily possible to distinguish between 
the normal and inverted mass hierarchy.

\subsubsection{\label{sec:0vbbQD}Quasi--degenerate Neutrinos}
The formulae for the mass states are given in Eq.\ (\ref{eq:massNH}). 
Ignoring $\eta$ and taking $U_{e3} = A \, \lambda$ one finds
\be
\frac{\meff}{m_0} \simeq \sqrt{c_\alpha^2 + 
\left( 2 - A^2 - (2 + A^2) \, c_{2 \alpha} + 2 \, A^2 \, c_\alpha \, 
c_{\alpha - 2 \beta}
\right) \, \lambda^2
} + {\cal O}(\lambda^3)~.
\ee
Interesting are the cases which correspond to $CP$ conservation, and 
which are obtained by setting $\alpha, \beta$ to $\pi/2,\pi$. 
They read up to ${\cal O}(\eta \lambda^2, \lambda^3)$:
\be \label{eq:rangeQD}
\frac{\meff}{m_0} \simeq 
\left\{ 
\bav  
1 - \eta & \alpha = \beta = 0 & \leftrightarrow & 
{\rm sign}(m_1) = {\rm sign}(m_2) = {\rm sign}(m_3) \\[0.3cm]
2 \, \lambda            & 2 \alpha = \beta = \pi & \leftrightarrow & 
{\rm sign}(m_1) = -{\rm sign}(m_2) = {\rm sign}(m_3)\\[0.3cm] 
2 \, \lambda            &  \alpha = \beta = \pi/2 & \leftrightarrow & 
{\rm sign}(m_1) = -{\rm sign}(m_2) = -{\rm sign}(m_3)\\[0.3cm] 
1 - \eta - 2 \, A^2 \lambda^2  & \alpha = 2 \beta = \pi & \leftrightarrow & 
{\rm sign}(m_1) = {\rm sign}(m_2) = -{\rm sign}(m_3)
\ea 
\right. 
\ee
As usual, for $U_{e3} = A \, \lambda^2$ and above, the dependence on $A$ 
(and thus also on $\beta$) drops and appears only to order 
$\lambda^4$. Noting that the minimal value of \meff{} is 
$2 \lambda \, m_0$, we can investigate if future experiments can distinguish 
between the normal and quasi--degenerate mass hierarchy \cite{PPR}. 
In analogy to the discussion leading to Eq.\ (\ref{eq:xiNHIH}), it follows 
from Eqs.\ (\ref{eq:meffNH1},\ref{eq:rangeQD})
\be
\xi \ls 2 \, \sqrt{\frac{2}{\eta}} \, 
\left(1 + 2 \, \lambda (1 \pm A^2) \right)~,
\ee
which will be easily possible. 
It is a bit more tricky to distinguish between the  quasi--degenerate
and the inverted hierarchies. The requirement for $\xi$ is 
\be
\xi \ls \sqrt{\frac{2}{\eta}} \, \lambda \, \left(
1 + \left(A^2 + \frac{1}{4} \right) \, 
\lambda^2 \right)~,
\ee
which is suppressed roughly by a factor $\lambda$ with respect to the 
limit on $\xi$ in order to distinguish the normal and quasi--degenerate 
mass hierarchy. Also in this aspect the parameter $\lambda$ shows up as 
a scaling factor.

\subsection{\label{sec:LBL}Long Baseline Oscillation Experiments}
There is another field of neutrino physics in which expansion in small 
parameters gives insight in the physics involved and which is therefore 
useful to study within our parameterization. 
These are the oscillation probabilities for long--baseline 
experiments \cite{CPLBL}. The determination of some 
currently unknown neutrino 
parameters, namely $U_{e3}$, the sign of \dma{} and the  
Dirac--like $CP$ violating phase are the purpose of such experiments. 
There are helpful expansions of the relevant oscillation probabilities 
in vacuum \cite{LBLexpa}. Here, we do not consider 
matter effects since they will not change our conclusions. 
Let us first comment on $CP$ violation. 
Using Eq.\ (\ref{eq:Pab}), 
one finds for the difference of the oscillation probabilities: 
\bea
\Delta P \equiv P(\nu_e \ra \nu_\mu) - 
P(\overline{\nu}_e \ra \overline{\nu}_\mu) = \\[0.3cm]
8 \, J_{CP} \, \left( 
\sin \frac{\D \Delta m^2_{31} \, L}{\D 2 \, E} 
-\sin \frac{\D \Delta m^2_{32} \, L}{\D 2 \, E} 
-\sin \frac{\D \Delta m^2_{21} \, L}{\D 2 \, E} 
\right)~,
\eea
which, using $\Delta m^2_{32} = \Delta m^2_{31} - \Delta m^2_{21}$, 
can easily be shown to vanish for two masses being equal.  
The invariant $J_{CP}$ 
was defined in Eq.\ (\ref{eq:JCPlep}). 
Since $\Delta m^2_{21}/\Delta m^2_{31} = \pm \lambda^2$, where the 
``+'' in case of normal ordering and the ``$-$'' for inverse ordering, 
we can expand the last equation:
\be
\Delta P \simeq \pm 4 \, J_{CP} \, \lambda^2 
\, \frac{\dma \, L}{2 \, E} \, 
\sin^2 \frac{\dma \, L}{4 \, E} 
+ {\cal O}(\lambda^3)~.
\ee
Thus, the $CP$ violating effects in realistic experiments 
are suppressed by another two 
orders of $\lambda$ in addition to the suppression present in $J_{CP}$.
If $n$ is the power of $\lambda$ in $U_{e3}$, then 
the total suppression is $\lambda^{2 + n}$. 

\noindent One can also consider the bare oscillation probability 
for the ``golden channel'', which 
is given by $\nu_e \ra \nu_\mu $ oscillations. 
Using the form of $P(\nu_e \ra \nu_\mu)$ as given, e.g., in 
\cite{reac}, one 
finds for the oscillation probability in case of $m = n = 1$: 
\be \D
P(\nu_e \ra \nu_\mu) \simeq 
2 \, A^2 \, \sin^2 \Delta_{32}  \, \lambda^2 
- 2 \, A \, \sin^2 \Delta_{32} \D 
\left( 2 \, A \, B + \cos (\delta \mp \Delta_{32}) \right) 
\, \lambda^3 + {\cal O}(\lambda^4)~. 
\ee
Here, the ``$-$'' sign is for neutrinos and the $+$ for antineutrinos. 
We defined $\Delta_{32} = (m_3^2 - m_2^2) \, L/4E 
$.   
The first term proportional to 
$\lambda^2$ is the term that probes $U_{e3}$ whereas the second 
term proportional to $\lambda^3$ is the one 
probing the $CP$ phase $\delta$.\\

\noindent As another example, assume $n = m = 2$. Then, the 
terms probing $U_{e3}$ and $CP$ violation will both 
be proportional to $\lambda^4$: 
\be 
P(\nu_e \ra \nu_\mu) \simeq \frac{\sin^2 \Delta_{32}}{2} \, 
\left( 
1 + 4 \, A^2 - 4 \, A \, \cos (\delta \mp \Delta_{32}) 
\right) \, \lambda^4 + {\cal O}(\lambda^6)~.
\ee
The parameter $B$ only appears at order $\lambda^6$, since there are no 
terms of order $\lambda^5$. For $m = 2$ and $n = 3$ it holds: 
\be
P(\nu_e \ra \nu_\mu) \simeq \frac{\sin^2 \Delta_{32}}{2} \, \lambda^4 -  
2 \, A \, \sin^2 \Delta_{32} \, \cos (\delta \mp \Delta_{32})  \, \lambda^5 
+ {\cal O}(\lambda^6)~.
\ee

\noindent A characteristic combination of the oscillation 
parameters that appears in the relevant probabilities is  
$\Delta m^2_{21}/\Delta m^2_{31}\, \sin 2 \theta_{12}$ \cite{LBLexpa}. 
Neglecting terms of order $\lambda^6$, we find for this 
parameter in our parametrization that 
\be
\frac{\Delta m^2_{21}}{\Delta m^2_{31} } \, \sin 2 \theta_{12} 
\simeq \pm \lambda^2 \, 
\left\{ \baz 
\left( 1 - 2 \, \lambda^2 + 2 \, (1 + A^2) \, \lambda^3 \right)
& \mbox{ for } n = 1 \\[0.3cm]
\left( 1 - 2 \, \lambda^2 + 2 \,  \lambda^3 \right)
& \mbox{ for } n = 2 \\[0.3cm]
\left( 1 - 2 \, \lambda^2 + 2 \, \lambda^3 \right)
& \mbox{ for } n = 3 
\ea 
\right.  ~,
\ee
where again the ``+'' in case of normal ordering 
and the ``$-$'' for inverse ordering. The difference between the 
cases $n=2$ and $n=3$ appears only at seventh order in $\lambda$. 
The characteristic parameter is therefore to order 
$\lambda^4$ independent on the precise form of the parametrization.

\section{\label{sec:concl}Conclusions}
The zeroth order approximation for neutrino mixing can be the bimaximal 
scheme with two maximal and one zero angle in the mixing matrix. 
It can be used as a reference matrix, whose corrections 
can be described in a similar manner as the Wolfenstein 
parametrization describes corrections to the unit matrix. 
Indeed, at least one of the angles in neutrino mixing 
is different from the extreme value corresponding to bimaximal mixing, 
namely the angle describing solar neutrino oscillations. 
To take this into account, a flexible 
parametrization of the neutrino mixing matrix was proposed 
in which the expansion parameter $\lambda \simeq 0.2$ is introduced 
to quantify this deviation from maximal mixing of solar neutrinos. 
It can also be used to quantify the possible 
deviation from zero $U_{e3}$ and 
maximal mixing of atmospheric neutrinos. The power of $\lambda$ to 
usefully describe these two latter aspects can be adjusted to future 
data. 
Depending on 
the power of $\lambda$, rather simple forms of the PMNS matrix are 
obtained, where the deviations from the ``bimaximal'' values 0, 
$\pm 1/2$ and $\pm 1/\sqrt{2}$  
are implied by $\lambda$. 
If $U_{\mu 3}$ and $U_{e3}$ are close to their maximally 
allowed values, $\lambda$ appears at first order in all 
elements of \PMNS. 
For values of 
$|U_{e3}| \ls 10^{-3}$ and $\sin^2 2\theta_{23} \gs 0.95$, the 
corrections become quadratic. 
The invariant measure for leptonic $CP$ violation is proportional to  
$\lambda^n$, where $n$ is the power of $\lambda$ in $U_{e3}$. 
One can interpret these corrections to the exact bimaximal 
mixing scheme in the same way as corrections to the unit matrix 
lead to the CKM matrix for the quark sector. 
\noindent 
Observing further that the ratio of the mass squared differences as measured 
in experiments is roughly $\lambda^2$, allows to study the 
form of the Majorana neutrino mass matrix $m_\nu$. 
Also here, the corrections to the extreme forms of $m_\nu$ in case 
of bimaximal mixing and extreme hierarchies are linear or quadratic in 
$\lambda$, depending on the precise values of $U_{e3}$, $U_{\mu 3}$ or the 
value of the smallest mass state.  
The $ee$ element of $m_\nu$ can be measured in experiments probing \onbb. 
Here, $\lambda$ appears as the scale factor of the typical values 
of \meff{} in the normal and inverted hierarchy. 
It also influences the maximal value of the 
uncertainty in the calculations of the nuclear 
matrix elements allowed to distinguish the normal, inverted or 
quasi--degenerated mass hierarchies. 
We furthermore commented on how our parametrization applies to 
realistic long--baseline oscillation experiments. 
Simple forms of the relevant oscillation probabilities are obtained.  
In particular, due to the small ratio of the two independent 
mass squared differences, effects of $CP$ violation 
are suppressed by another two orders of $\lambda$.

\vspace{0.5cm}
\begin{center}
{\bf Acknowledgments}
\end{center}
I thank S.T.\ Petcov for encouragement and 
helpful comments and am grateful to P.H.~Frampton for 
discussions. 
This work was supported by 
the EC network HPRN-CT-2000-00152.


\begin{thebibliography}{99} 
\bibitem{reviews}For recent reviews see, e.g.:
M.~C.~Gonzalez-Garcia and Y.~Nir, 
Rev.\ Mod.\ Phys.\  {\bf 75}, 345 (2003); 
Z.~z.~Xing, hep-ph/0307359; 
V.~Barger, D.~Marfatia and K.~Whisnant, 
Int.\ J.\ Mod.\ Phys.\ E {\bf 12}, 569 (2003). 

\bibitem{KamLAND}K.~Eguchi {\it et al.}  [KamLAND Collaboration], 
Phys.\ Rev.\ Lett.\  {\bf 90}, 021802 (2003). 


\bibitem{sol}B.~T.~Cleveland {\it et al.}, 
Astrophys.\ J.\  {\bf 496}, 505 (1998); 
Y.~Fukuda {\it et al.}  [Kamiokande Collaboration],
Phys.\ Rev.\ Lett.\  {\bf 77}, 1683 (1996); 
V.~N.~Gavrin  [SAGE Collaboration],
Nucl.\ Phys.\ Proc.\ Suppl.\  {\bf 91}, 36 (2001); 
W.~Hampel {\it et al.}  [GALLEX Collaboration], 
Phys.\ Lett.\ B {\bf 447}, 127 (1999); 
M.~Altmann {\it et al.}  [GNO Collaboration],
Phys.\ Lett.\ B {\bf 490}, 16 (2000); 
Q.R. Ahmad \textit{et al.}, [SNO Collaboration],  Phys. Rev. Lett.\ 
{\bf 89} (2002) 011302 and 011301.  
S.~Fukuda {\it et al.}  [Super-Kamiokande Collaboration], 
Phys.\ Lett.\ B {\bf 539}, 179 (2002). 

\bibitem{SNOII}S.~N.~Ahmed {\it et al.}  [SNO Collaboration], 
nucl-ex/0309004.

\bibitem{SNOana1}A.~Bandyopadhyay, {\it et al.}, hep-ph/0309174.

\bibitem{SNOana2}
A.~B.~Balantekin and H.~Yuksel, hep-ph/0309079;
G.~L.~Fogli, {\it et al.}, hep-ph/0309100; 
M.~Maltoni {\it et al.}, hep-ph/0309130; 
P.~Aliani, {\it et al.}, hep-ph/0309156;
P.~C.~de Holanda and A.~Y.~Smirnov, hep-ph/0309299.



\bibitem{SKatm}Y.~Fukuda {\it et al.}  [Super-Kamiokande Collaboration],
Phys.\ Rev.\ Lett.\  {\bf 82}, 2644 (1999). 

\bibitem{K2K}M.~H.~Ahn {\it et al.}  [K2K Collaboration],
Phys.\ Rev.\ Lett.\  {\bf 90}, 041801 (2003).

\bibitem{reactor}F.~Boehm {\it et al.},
Phys.\ Rev.\ D {\bf 64}, 112001 (2001);
M.~Apollonio {\it et al.},
Eur.\ Phys.\ J.\ C {\bf 27}, 331 (2003)


\bibitem{wolf}L.~Wolfenstein,
Phys.\ Rev.\ Lett.\  {\bf 51}, 1945 (1983).

\bibitem{kaus}P.~Kaus and S.~Meshkov,
AIP Conf.\ Proc.\  {\bf 672}, 117 (2003) (hep-ph/0211338).

\bibitem{zing}Z.~z.~Xing, 
J.\ Phys.\ G {\bf 29}, 2227 (2003). 

\bibitem{zee}A.~Zee, Phys.\ Rev.\ D {\bf 68}, 093002 (2003).


\bibitem{PDG}K.~Hagiwara, {\it et al.}  [Particle Data Group Collaboration],
Phys.\ Rev.\ D {\bf 66}, 010001 (2002).

\bibitem{JCP}C.~Jarlskog, 
Phys.\ Rev.\ Lett.\  {\bf 55}, 1039 (1985).

\bibitem{PMNS}B. Pontecorvo, Zh. Eksp. Teor. Fiz.\ {\bf 33}, 549 (1957) 
and {\bf 34}, 247 (1958); 
Z. Maki, M. Nakagawa and S. Sakata, Prog. Theor. Phys.\ {\bf 28}, 870 (1962).

\bibitem{Majpha}
S.~M.~Bilenky, J.~Hosek and S.~T.~Petcov,
Phys.\ Lett.\ B {\bf 94}, 495 (1980); 
J.~Schechter and J.~W.~Valle, 
Phys.\ Rev.\ D {\bf 22}, 2227 (1980);
M.~Doi, {\it et al.}, Phys.\ Lett.\ B {\bf 102}, 323 (1981); 



\bibitem{CPLBL}M.~Lindner, 
invited talk at 20th International Conference on 
Neutrino Physics and Astrophysics (Neutrino 2002), Munich, 
Germany, 25-30 May 2002, hep-ph/0210377.


\bibitem{carlos}M.~C.~Gonzalez-Garcia and C.~Pena-Garay, 
Phys.\ Rev.\ D {\bf 68}, 093003 (2003).


\bibitem{SKaachen}Y. Hayato, talk Presented at 
International Europhysics Conference on High-Energy Physics 
(HEP 2003), Aachen, Germany, 17-23 Jul 2003; 
{\tt http://eps2003.physik.rwth-aachen.de/transparencies/07/index.php}

\bibitem{LBL}Y.~Itow {\it et al.}, hep-ex/0106019;
D.~Ayres {\it et al.}, hep-ex/0210005.


\bibitem{reac}P.~Huber, {\it et al.},  
Nucl.\ Phys.\ B {\bf 665}, 487 (2003). 


\bibitem{nufac}M.~Apollonio {\it et al.}, hep-ph/0210192.


\bibitem{low}S.~Sch\"onert, 
Nucl.\ Phys.\ Proc.\ Suppl.\  {\bf 118}, 62 (2003),  
talk given at 20th International Conference on Neutrino Physics 
and Astrophysics (Neutrino 2002), Munich, Germany, 25-30 May 2002. 



\bibitem{STP102}P.~I.~Krastev and S.~T.~Petcov, 
Phys.\ Lett.\ B {\bf 205} (1988) 84.

\bibitem{bimax}
F.~Vissani, hep-ph/9708483; 
V.~D.~Barger, S.~Pakvasa, T.~J.~Weiler and K.~Whisnant, 
Phys.\ Lett.\ B {\bf 437}, 107 (1998); 
A.~J.~Baltz, A.~S.~Goldhaber and M.~Goldhaber, 
Phys.\ Rev.\ Lett.\  {\bf 81}, 5730 (1998); 
H.~Georgi and S.~L.~Glashow,
Phys.\ Rev.\ D {\bf 61}, 097301 (2000). 


\bibitem{devbimax1}C.~Giunti and M.~Tanimoto, 
Phys.\ Rev.\ D {\bf 66}, 053013 (2002); 
Phys.\ Rev.\ D {\bf 66}, 113006 (2002).
 
\bibitem{radcor}See, e.g.,  
J.~A.~Casas, {\it et al.}, 
Nucl.\ Phys.\ B {\bf 573}, 652 (2000); 
P.~H.~Chankowski and S.~Pokorski, 
Int.\ J.\ Mod.\ Phys.\ A {\bf 17}, 575 (2002); 
S.~Antusch, {\it et al.}, hep-ph/0305273 and 
references therein.



\bibitem{devbimax2}
T.~Miura, T.~Shindou and E.~Takasugi, hep-ph/0308109.



\bibitem{smir}M.~C.~Gonzalez-Garcia, {\it et al.}, 
Phys.\ Rev.\ D {\bf 63}, 013007 (2001). 

\bibitem{lelmlt}S.~T.~Petcov, 
Phys.\ Lett.\ B {\bf 110} (1982) 245.


\bibitem{seesaw}M. Gell--Mann, P. Ramond, 
and R. Slansky in {\it Supergravity},
p. 315, edited by F. Nieuwenhuizen 
and D. Friedman, North Holland,
Amsterdam, 1979;
T. Yanagida, Proc. of the 
{\it Workshop on Unified Theories and the Baryon
Number of the Universe}, edited by 
O. Sawada and A. Sugamoto, KEK, Japan 1979;
R.N. Mohapatra, G. Senjanovic, \Jo{\PRL}{44}{912}{1980}.

\bibitem{Majins}S.~M.~Bilenky, J.~Hosek and S.~T.~Petcov 
in \cite{Majpha}; 
J.~Schechter and J.~W.~Valle, 
Phys.\ Rev.\ D {\bf 23}, 1666 (1981); 
S.~M.~Bilenky, N.~P.~Nedelcheva and S.~T.~Petcov, 
Nucl.\ Phys.\ B {\bf 247}, 61 (1984).


\bibitem{ichL}W. Rodejohann, 
Phys.\ Rev.\ D {\bf 62}, 013011 (2000); 
J.\ Phys.\ G {\bf 28}, 1477 (2002); 
A.~de Gouvea, B.~Kayser and R.~N.~Mohapatra,
Phys.\ Rev.\ D {\bf 67} (2003) 053004. 


\bibitem{frig}M.~Frigerio and A.~Y.~Smirnov,
Nucl.\ Phys.\ B {\bf 640}, 233 (2002); 
Phys.\ Rev.\ D {\bf 67}, 013007 (2003); see also the first two 
references in \cite{ichL}.


\bibitem{0vbb}E.g., 
F.~Vissani, JHEP {\bf 9906}, 022 (1999); 
S.~M.~Bilenky, {\it et al.}, Phys.\ Lett.\ B {\bf 465}, 193 (1999); 
Phys.\ Rev.\ D {\bf 64}, 053010 (2001); 
H.~V.~Klapdor-Kleingrothaus, H.~P\"as and A.~Y.~Smirnov, 
Phys.\ Rev.\ D {\bf 63}, 073005 (2001); 
W.~Rodejohann, 
Nucl.\ Phys.\ B {\bf 597}, 110 (2001); 
F.~Feruglio, A.~Strumia and F.~Vissani,
Nucl.\ Phys.\ B {\bf 637}, 345 (2002)
[Addendum-ibid.\ B {\bf 659}, 359 (2003)]; 
V.~Barger, {\it et al.}, 
Phys.\ Lett.\ B {\bf 532}, 15 (2002); 
F.~R.~Joaquim, 
Phys.\ Rev.\ D {\bf 68}, 033019 (2003); 
see also S.~Pascoli, S.~T.~Petcov, hep-ph/0308034 for a rather 
complete list of references.




\bibitem{NME}S.~M.~Bilenky and J.~A.~Grifols,
Phys.\ Lett.\ B {\bf 550}, 154 (2002); 
A.~Faessler, {\it et al.}, 
Phys.\ Rev.\ C {\bf 68}, 044302 (2003);  
O. Civitarese, J. Suhonen, nucl-th/0208005 and 
references therein. 

\bibitem{carlos2}H.~Murayama and C.~Pena-Garay, hep-ph/0309114.



\bibitem{PPR}S.~Pascoli, S.~T.~Petcov and W.~Rodejohann, 
Phys.\ Lett.\ B {\bf 549} (2002) 177; 
Phys.\ Lett.\ B {\bf 558} (2003) 141. 


\bibitem{LBLexpa}A.~Cervera, {\it et al.}, 
Nucl.\ Phys.\ B {\bf 579}, 17 (2000)
[Erratum-ibid.\ B {\bf 593}, 731 (2001)]; 
M.~Freund, Phys.\ Rev.\ D {\bf 64}, 053003 (2001); 
V.~Barger, D.~Marfatia and K.~Whisnant, 
Phys.\ Rev.\ D {\bf 65}, 073023 (2002); see also 
\cite{CPLBL}. 

\end{thebibliography}
\end{document}